\documentclass[pra,twocolumn,amsmath,amssymb,groupedaddress,longbibliography]{revtex4-2}
\usepackage{bbm}
\usepackage{mathrsfs}
\usepackage{amsmath}
\usepackage{amsfonts}
\usepackage[pdfstartview=FitH,colorlinks=true,citecolor=blue,anchorcolor=blue]{hyperref}
\usepackage{graphicx}
\usepackage{subfigure}
\usepackage{dcolumn}
\usepackage{bm}
\usepackage{color}
\usepackage{natbib}
\usepackage{amssymb}
\usepackage{xcolor}
\usepackage{braket}
\usepackage{soul}

\begin{document}

\title{Nonlinear topological pumping of edge solitons}
\author{Xinrui You$^{1,2}$}

\author{Liaoyuan Xiao$^{1,2}$}

\author{Yongguan Ke$^{2}$} \email{keyg@szu.edu.cn}

\author{Chaohong Lee$^{2,3}$}

\affiliation{$^{1}$Laboratory of Quantum Engineering and Quantum Metrology, School of Physics and Astronomy, Sun Yat-Sen University (Zhuhai Campus), Zhuhai 519082, China}

\affiliation{$^{2}$Institute of Quantum Precision Measurement, State Key Laboratory of Radio Frequency Heterogeneous Integration, College of Physics and Optoelectronic Engineering, Shenzhen University, Shenzhen 518060, China}
\affiliation{$^{3}$Quantum Science Center of Guangdong-Hong Kong-Macao Greater Bay Area (Guangdong), Shenzhen 518045, China}

\date{\today}

\begin{abstract}
We study how nonlinear strength affects topological pumping of edge solitons by using  nonlinear Gross–Pitaevskii equation.
For weak nonlinear strength, the introduction of nonlinearity breaks the symmetry of the energy spectrum, which makes the topological pumping from the left edge to the right edges differ from the inverse process.
For moderate nonlinear strength, self-crossing structures appear in the spectrum, the right-to-left adiabatic pumping channel is destroyed, and only left-to-right topological pumping can be achieved under slow modulation. 
As the nonlinear strength further inreases, although left-to-right topological pumping in one pumping cycle also breaks down, we find that a thin soliton  which is located in a single left edge can be mixed with the bulk soliton, and hybridized topological pumping of edge and bulk solitons can be realized after multiple pumping cycles.
For stronger nonlinear strength, edge solitons are self-trapped and all topological pumping channels are shut down. 
Our work could trigger further studies of the interplay between nonlinearity and topology.
\end{abstract}

\maketitle


\section{Introduction} \label{INTRODUCTION}
Topological pumping, particle transport protected by topological properties of spatiotemporal modulation, has garnered intense research interest in recent years~\cite{citro2023thouless}. 
Of particular interest are Thouless pumping~\cite{PhysRevB.27.6083} and topological pumping of edge states~\cite{PhysRevLett.109.106402,cheng2022asymmetric}. 
In Thouless pumping, noninteracting particles uniformly and adiabatically sweeping an energy band will be transferred by unit cells per pumping cycle, where the displacement is determined by the Chern number of the band~\cite{PhysRevB.47.1651,RevModPhys.82.1959}. 
In topological pumping of edge states, the edge states can be transferred from one end to the other end through the edge-state channels in the band gap.
These topological pumps have been experimentally realized in many systems, including ultracold atoms~\cite{lohse2016thouless,nakajima2016topological,PhysRevLett.111.026802} in optical lattices, photonic waveguides~\cite{PhysRevB.91.064201,KeLPR2016,cerjan2020thouless,sun2022non}, and spin systems~\cite{PhysRevLett.120.120501}.
The robustness of topological pumping against disorder and imperfection is often harnessed for implementing quantum information processes~\cite{alicea2011non,nayak2008non,lang2017topological}, such as topological quantum state transfer~\cite{bello2016long,dlaska2017robust,PhysRevA.98.012331,longhi2019landau,PhysRevB.99.155150,
Qi:20,PhysRevResearch.2.033475,PhysRevA.103.052409}, topological quantum interference~\cite{tambasco2018quantum,
PhysRevLett.120.047702,bardarson2013quantum,PhysRevA.101.052323}, and topological quantum gates~\cite{He:19,PhysRevB.100.045414,He:20,chao2021novel,narozniak2021quantum}.

The interplay between topology and particle-particle interaction or nonlinearity gives rise to novel phenomena. 
For few strongly interacting particles, topological pumping of bound states~\cite{PhysRevA.95.063630,van2019topological,PhysRevResearch.5.013020,PhysRevA.101.023620,huang2024topological} and topologically resonant tunnelings~\cite{PhysRevA.95.063630} have been theoretically predicted and experimentally realized in cold atomic systems~\cite{walter2023quantization,ke2023topological}.   
For many weakly interacting particles, particle-particle interaction can be treated in the mean-field approximation~\cite{PhysRevB.93.020502,gulevich2017exploring,leykam2016edge,bleu2016interacting,bleu2017photonic,LIU2017183}, and motions of particles can be described by nonlinear Gross–Pitaevskii (GP) equation~\cite{gross1961structure,
pitaevskii1961vortex}, where solitons are kinds of well-known solitons.
Recently, topological pumping of soliton has been explored theoretically~\cite{PhysRevLett.128.113901,mostaan2022quantized,PhysRevLett.128.154101,PhysRevLett.129.183901,Tuloup_2023,16-20230740,PhysRevResearch.6.L042010} and experimentally~\cite{jurgensen2021quantized,
jurgensen2023quantized}, where soliton may exhibit quantized and fractional transport in analogy to Thouless pump. 
The (fractional) quantized displacement of the soliton is governed by (average) Chern numbers of single-particle energy bands, depending on the non-linearity strength~\cite{PhysRevLett.128.154101,
jurgensen2023quantized}. 
Inheriting from the bulk-edge correspondence in noninteracting topological systems~\cite{PhysRevB.94.041102}, there exist topological edge solitons when adding nonlinearity.
However, it remains an open question how the topological pumping of an edge soliton changes with nonlinear strength, which involves a complex relationship between nonlinear dynamics and topological phase. 
%

\begin{figure*}[htpb]
    \centering
    \includegraphics[width=\linewidth]{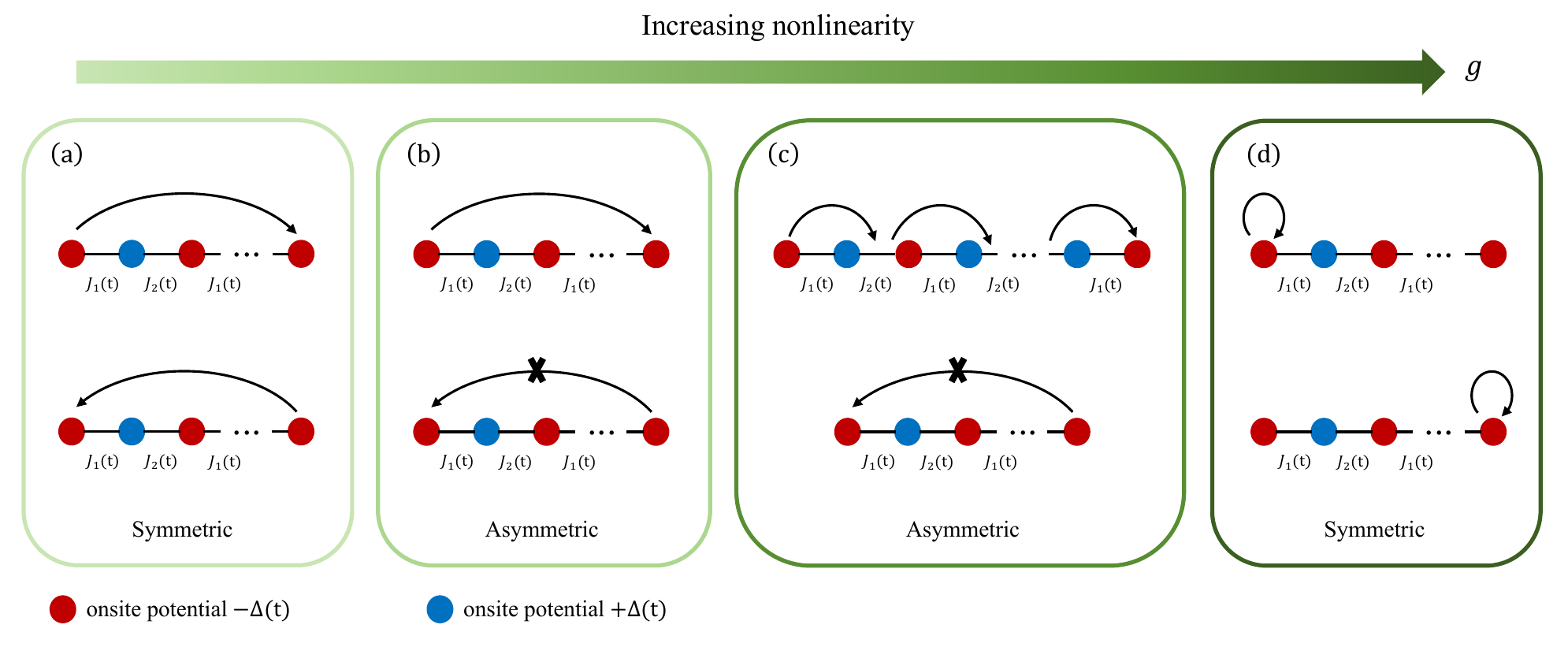}
    \caption{Schematics for pumping dynamics of edge solitons in modulated superlattice as nonlinear strength ($g$) increases. 
    $J_{1}(t)=-[J-\delta \cos(\omega t)]$, $J_{2}(t)=-[J+\delta \cos(\omega t)]$, and $\Delta(t)=\Delta_0 \sin(\omega t)$ denote the inter-unit cell hopping, intra-unit cell hopping and staggered on-site potential, respectively.
    (a) In the linear or weakly nonlinear case, edge solitons can be transferred from the left to the right edges in one pumping cycle, and vice versa. 
    (b) In the moderate nonlinear case, one can only realize unidirectional transport from the left to the right edges in one pumping cycle, and the reversed process breaks down. 
    (c) When the nonlinear strength is further enhanced to a certain extent, the edge soliton also cannot be transport from the left to the right edges in one pumping cycle. However, after multiple pumping cycles, edge soliton can be transferred from the left to the right edges via hybridization of Thouless pumping of bulk soliton.
    (d) In the strong nonlinear case, edge solitons are self-trapped.}
    \label{fig:enter-label1}
\end{figure*}

In this paper, we investigate topological transport of edge soliton states by considering weakly interacting bosonic atoms in a periodically modulated optical lattice. 
The transport of edge solitons behaves quite differently in weak, medium, and strong nonlinear regions; see the schematics in Fig.~\ref{fig:enter-label1}. 
In the weak nonlinearity region, the edge soliton behavior closely resembles that of edge-state transport in linear systems. 
The edge soliton evolves adiabatically along the instantaneous nonlinear eigenstates, enabling successful transport from the left edge to the right edge or vice versa [Fig.~\ref{fig:enter-label1}(a)].
At moderate nonlinear strength, the edge soliton can be transferred from the left edge to the right edge, while the inverse  transport fails, regardless of how slow the drive is [Fig.~\ref{fig:enter-label1}(b)]. 
This is because  the energy bifurcation in the right-to-left transport channel makes the instantaneous evolved states cannot follow the instantaneous eigenstates. 
As nonlinearity increases further, we identify a mixture of topological pumping of edge states and Thouless pumping, that is, the most localized soliton at the left edge is first transferred to a bulk soliton and then the bulk soliton undergoes Thouless pumping for several cycles and finally becomes a soliton at the right edge [Fig.~\ref{fig:enter-label1}(c)]. 
In the strong nonlinearity region, all edge solitons including solitons at the edge and sub-edge become self-trapped [Fig.~\ref{fig:enter-label1}(d)]. 
The sub-edge solitons come from the energy bifurcation of edge-soliton solutions.

The rest of paper is organized as follows. In Sec.~\ref{RICE-MELE MODEL}, we describe the nonlinear Rice-Mele (RM) model and numerical methods for dynamically evolving edge solitons and nonlinear energy spectra. 
In Sec.~\ref{NONLINEAR EDGE STATE TOPOLOGICAL TRANSPORT}, we investigate edge soliton transport under varying nonlinear strengths and interpret the results using the nonlinear energy spectrum.
In Sec.~\ref{CONCLUSION}, we give a brief summary and a discussion.

\section{MODEL AND METHOD} \label{RICE-MELE MODEL}
Ultracold atomic system serves as an excellent platform to study interplay between topology and interaction, in which different structures of optical lattices can be designed and dynamically driven and particle-particle interaction can be precisely tuned by Feshbach resonance techniques~\cite{bloch2005ultracold,PhysRevLett.110.215301,PhysRevB.92.245409,PhysRevLett.115.095302,PhysRevB.94.235139,goldman2016topological,tai2017microscopy,lindner2017universal}. 
Thouless pumps of noninteracting bosonic and fermionic atoms have been realized in modulated superlattice described by RM model~\cite{PhysRevLett.111.026802,nakajima2016topological,
lohse2016thouless}. 
Ramping particle-particle interaction, a recent experiment has observe Thouless pumping of bound states in which two particles move unidirectionally  as a whole, and topologically resonant tunneling in which particles are shifted one bye one~\cite{walter2023quantization,
ke2023topological}.
Owing to the well-control setup of ultracold atoms, we consider weakly interacting bosonic gas in periodically driven superlattice, which obeys an interacting RM model~\cite{PhysRevA.95.063630},
    \begin{equation}\label{eq01}
    \begin{aligned}
    \hat{H}=&\sum_{j}J_{j}(\hat{a}^{\dagger}_{j} \hat{a}_{j+1}+H.c.)+\sum_{j}[\Delta_{j} \hat{n}_{j}+\frac{g}{2}\hat{n}_{j}(\hat{n}_{j}-1)].
    \end{aligned}
    \end{equation}
Here, $\hat{a}^{\dagger}_{j}$ ($\hat{a}_{j}$) are bosonic creation (annihilation) operators at the $j$th site, $\hat{n}_{j}={a}^{\dagger}_{j} {a}_{j}$ is the density operator,  $g$ is the strength of onsite interaction, $J_{j}=-[J+(-1)^{j}\delta \cos(\omega t)]$ and $\Delta_{j}=(-1)^{j} \Delta \sin(\omega t)$ represent the modulated nearest-neighbor hopping amplitude and on-site energy with pumping cycle $T=2 \pi/ \omega$, respectively. For simplicity, we set $\hbar=1$ and other parameters $\{\delta, \Delta, g, \hbar \omega \}$ in units of the hopping constant $J$.

\begin{figure*}[!htp]
    \centering
    \includegraphics[width=0.8\linewidth]{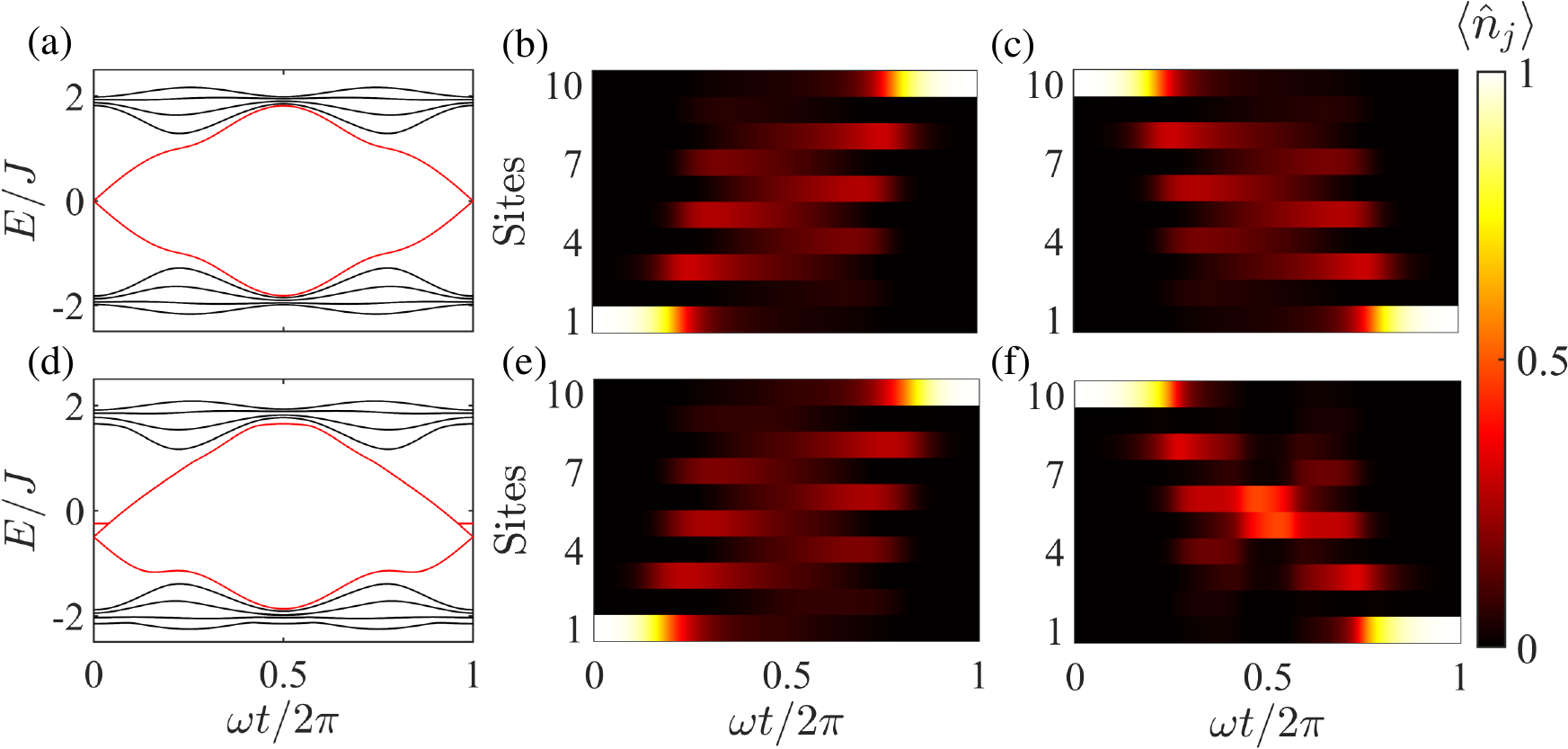}
    \caption{Top panel: Energy spectrum and symmetric topological pumping of  edge states in linear Rice-Mele model. 
    (a) Linear energy spectrum over one cycle. 
    Red lines denote localized edge states. 
   Time evolution of density distribution for initial state prepared in the (b) left edge and (c) right edge.
    Bottom panel: Nonlinear energy spectrum and asymmetric topological pumping of edge solitons in the weak nonlinear case $g=0.5$.
    (d-f) are similar to (a-c) but with nonzero nonlinear strength and initial states. 
    The parameters are chosen as $\delta=0.9$, $\Delta_{0}=J=1$, $T=5000$, and the system size is $N=10$. }
    \label{fig:enter-label2}
\end{figure*}

We adopt the mean-field approximation so that many-body interacting problems are reduced to nonlinear GP equation,
    \begin{equation}\label{eq1}    
    i\dot{\psi}_{j}(t)=\sum_{j}H_{j}^{\mathrm{lin}}(t)\psi_{j}(t)-g|\psi_{j}(t)|^{2}\psi_{j}(t).   
    \end{equation}
 Here, $\psi_{j}$ is the amplitude of the wave function at site $j$ and time $t$, $H^{\mathrm{lin}}$ is the linear time-dependent RM Hamiltonian. 
 In the following, we only consider attractive interaction with $g>0$, and $\dot{\psi}_{j}$ represents the time derivative of $\psi_{j}$. 
 %
 The linear RM model~\cite{PhysRevLett.49.1455} is a prototypical model to illustrate the Thouless pump in a lattice with two sites per unit cell,
    \begin{equation}\label{eq2}
    \begin{aligned}
    H_{j}^{\mathrm{lin}}(t)=&\sum_{j}J_{j}(\hat{a}^{\dagger}_{j} \hat{a}_{j+1}+H.c.)+\Delta_{j} \hat{n}_{j}.
    \end{aligned}
    \end{equation}

 We will focus on the time evolution of edge solitons by solving Eq.~\eqref{eq1}, which can be achieved with the 4th order Runge Kutta method.
To verify whether the evolved states adiabatically follow the instantaneous nonlinear eigenstates, 
we first need to numerically calculate instantaneous eigenstates. 
Assuming $\psi_{j}(t) \rightarrow e^{-i\lambda t}\psi_{j}$ where $\lambda$ is the nonlinear eigenvalue, Eq.~\eqref{eq1} takes the following form,
\begin{equation}\label{eq3}
\sum_{j}H_{j}^{\mathrm{lin}}\psi_{j}-g|\psi_{j}|^{2}\psi_{j}-\lambda\psi_{j}=0.
\end{equation}
 Without loss of generality, we normalize wave functions ($\sum_{j}|\psi_{j}|^{2}=1$) and use Newton-Jacobi iterative method to solve the set of $N+1$ equations (with $N$ being the system size), which can give instantaneous stationary states and nonlinear eigenvalues; see details in Appendix~\ref{app1}. 
 
The success of obtaining a stable solution mainly depends on the initial guess, so the choice of the initial guess is important.
For weak nonlinear strength,  linear eigenvalues and eigenstates are used as an initial guess. 
However,  for large nonlinear strength, the choice of initial guess is more complex.
For a given nonlinear $g$ and state-dependent Hamiltonian $H$ at a given time $t$, the iterative process of obtaining $|\psi_{j}^{g}(t\pm \Delta t)\rangle$  from the state $|\psi_{j}^{g}(t) \rangle$  is as follows.

1. Nonlinear eigenvalue $\lambda_{t}^{g-\Delta g}$ and eigenstate $|\psi_{j}^{g-\Delta g}(t) \rangle$ at lower nonlinear strength are used as initial guesses to obtain the nonlinear eigenvalue  $\lambda_t^{g}$ and eigenstate $|\psi_{j}^{g}(t) \rangle$ at higher nonlinear strength.

2. The nonlinear eigenvalue and eigenstate at time $t$ are used as initial guess to obtain the nonlinear eigenvalue $\lambda_{t+\Delta t}^{g}$ ($\lambda_{t-\Delta t}^{g}$)  and eigenstate $|\psi_{j}^{g}(t+\Delta t) \rangle$ ($|\psi_{j}^{g}(t-\Delta t) \rangle$) at the late (previous) time.

Note that sometimes a nonlinear eigenstate and its corresponding eigenvalue as a set of initial guesses may not be able to obtain the instantaneous stable state.
In this case, we need more initial guesses. 
We can optimize it to take more eigenstates around the a certain energy to form a set of initial guesses.
Because the nonlinear eigenvalues of the two adjacent moments are relatively close, it is impossible to mutate.
In order to ensure the accuracy of the results, we add a restriction that the difference between the nonlinear eigenvalues of adjacent time steps should be less than $0.1$.

\section{Pumping dynamics of edge solitons}\label{NONLINEAR EDGE STATE TOPOLOGICAL TRANSPORT}
In this section we investigate the transport of edge solitons for different nonlinear strengths in the nonlinear RM model; see Fig.~\ref{fig:enter-label1}. 
%
%
In subsection~\ref{SubA}, we show that the symmetric topological pumping in linear systems turns to asymmetric one as tiny nonlinearity is ramped up, because nonlinearity breaks the symmetric structure of the edge-state transport channels.
In subsection~\ref{A}, for intermediate nonlinear strength, we show that the edge soliton can only be transported from left edge to right edge in one pumping cycle, and reversed transport fails due to the presence of self-crossing in nonlinear topological bands.
In subsection~\ref{hybridized}, as the nonlinear strength increases further, we show a hybridized topological pumping as a mix transport of edge soliton and bulk soliton from left to right, while the reversed process is also forbidden. 
In subsection~\ref{selftrap}, under strong nonlinearity, we will show edge and sub-edge solitons are self trapped.

\subsection{Dual-channel topological pumping} \label{SubA}
Before taking the nonlinearity into account, we first consider the linear case, Eq.~\eqref{eq1} reduces to the \text{Schrödinger} equation. 
According to the bulk-edge correspondence, the lower band has a Chern number of $1$, indicating two edge states in the bulk gap.
In Fig.~\ref{fig:enter-label1}(a), we show the energy spectrum of the RM model is symmetric with respect to zero energy. 
The parameters are chosen as $J= \Delta_0= 1$ and $\delta = 0.9$, with $L=5$ unit cells ($N = 10$ sites), and we impose open boundary conditions with $\psi_{\text{-} 1} =\psi_{N+1} = 0$.
In the bulk gap, there are two topological edge states marked in red which provide reversible channels for topological pumping of edge states from the left end to the right end or vise versa.

Because the energies of the two channels are symmetric about zero, the topological pumping from the left to the right edges is symmetric with the reversed process; see the time evolution of density distribution in Fig.~\ref{fig:enter-label1}(b) and (c). Obviously, the initial state prepared at one edge can both be transferred to the other edge successfully, meaning the symmetric transports can be implemented in the standard RM model.

 However, even when tiny nonlinearity is involved, the picture becomes totally different. 
 In the weak nonlinear region ($g=0.5$), we observe that the nonlinear spectrum has already slightly changed [Fig.~\ref{fig:enter-label2}(d)], in which two edge channels become asymmetric. 
Because the nonlinearity breaks the symmetric energy spectrum, the previous symmetric pumping in the linear case becomes slightly asymmetric; see Figs.~\ref{fig:enter-label2}(e) and (f).
However, edge soliton can be successfully transported from the left edge to the right edge, or vice versa. 
In this case, both left-to-right and right-to-left channels support topological pumping.
By calculating instantaneous eigenstate $|\psi_{I}(t)\rangle$ and the time evolution state $|\Psi_{E}(t)\rangle$ at time $t$ by solving Eq.~\eqref{eq1}, the probability of the instantaneous state projected into the eigenstates can be given by
\begin{equation}
    P(t)=|\langle \psi_{I}(t) | \psi_{E}(t) \rangle|^2.
\end{equation}
We find that $P(t)\approx 1$ in one pumping cycle in the linear (weak nonlinear) case, indicating that the edge (soliton) states can evolve adiabatically along the instantaneous nonlinear eigenstate and achieve successful topological pumping.

\begin{figure}[htpb]
    \centering
    \includegraphics[width=1.03\linewidth]{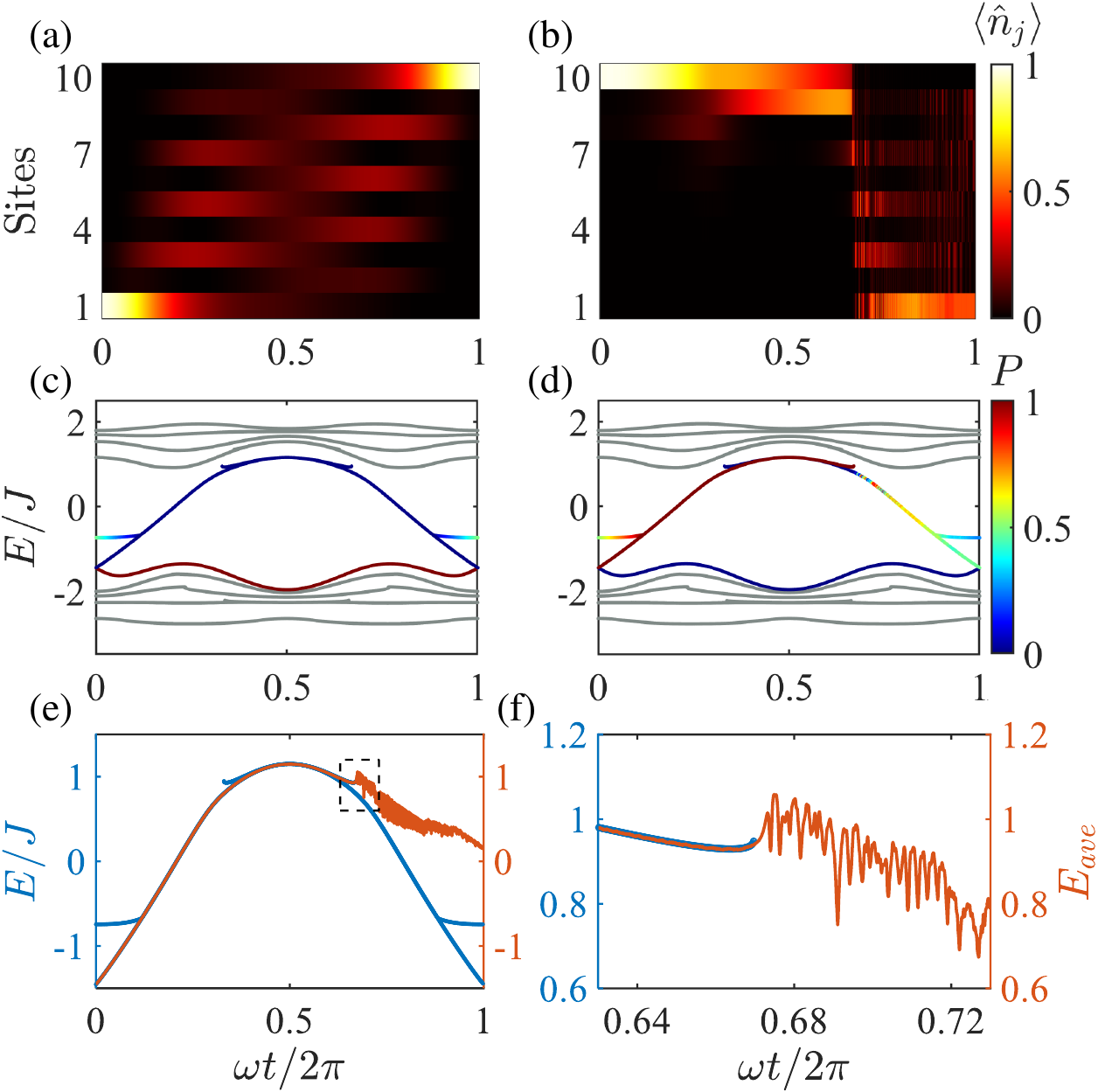}
    \caption{Time evolution of edge solitons, nonlinear energy spectrum, energy of instantaneous states. Time evolution of density distribution for the edge solitons prepared in the (a) left edge and (b) right edge. 
    (c, d) Nonlinear energy spectrum with the colors denote the projection probability of instantaneous evolved states in (a,b) onto the instantaneous eigenstates, respectively.
    (e) Energy as function of time. $E$ on the left side denotes energy of instantaneous soliton and $E_{ave}$ on the right side denotes mean energy of evolved states. (f) Enlarged region marked by the dashed rectangle in (e). The parameters are the same as those in Fig.~\ref{fig:enter-label2} except for $g=1.5$.}
    \label{fig:enter-label3}
\end{figure}

\subsection{Single-channel topological pumping}\label{A}
As nonlinear strength increases, in the intermediate nonlinear region ($g=1.5$), only topological pumping from the left edge to the right edge can be implemented, while reverse transport from the right edge to the left edge is forbidden; see Figs.~\ref{fig:enter-label3}(a) and (b), respectively. 
The other parameters are chosen as those in Fig.~\ref{fig:enter-label2}.
To understand the mechanics, we calculate the energy spectrum as a function of time in one pumping cycle; see Figs.~\ref{fig:enter-label3}(c) and (d).
Because we focus on the nonlinear edge soliton states instead of the nonlinear bulk states, the energies of nonlinear bulk states are marked as dark lines.
Although the energies of the lower and upper bands are more or less symmetric about $0$ for weak nonlinear strength [Fig.~\ref{fig:enter-label2}(d)], they apparently become more asymmetric for intermediate nonlinear strength.
The colors in the energies of the edge soliton states denote the projection probability of instantaneous evolved states onto the edge soliton eigenstates, and the difference between Figs.~\ref{fig:enter-label3}(c) and (d) is that the initial states are edge soliton states located at the left and right edges, respectively.  
We can clearly see that the projection probabilities in the lower left-to-right channel are quite close to $1$.
It means that the initial edge soliton at the left edge can adiabatically follow the lower left-to-right channel [Fig.~\ref{fig:enter-label3}(c)].
However, there are energy bifurcations in the upper right-to-left channel.
The initial edge soliton at the right edge can follow the right-to-left channel until it reaches the ‘dead end’ around $t \approx 0.67T$~\cite{Tuloup_2023}, beyond which there is no dynamically stable edge soliton to adiabatically follow [Fig.~\ref{fig:enter-label3}(d)].
Under the same parameters of the linear RM model as those in Fig.~\ref{fig:enter-label2}, we can find that the critical value of the nonlinearity strength is about $g_{c,1}=1.2$, above which energy bifurcations make the topological pumping change from dual channels to single channel.

We also calculate the expectation value of the time-dependent nonlinear Hamiltonian,
\begin{equation}
 E_{ave}(t)=\langle \Psi_{E}(t) |\hat{H}(t)| \Psi_{E}(t) \rangle,
\end{equation}
where the wave function $|\Psi_{E}(t)\rangle$ at time $t$ is obtained by solving equation Eq.~\eqref{eq1}. 
While $E_{ave}(t)$ coincides with the eigenvalues of the right-to-left channel before the time of ‘dead end', after which $E_{ave}(t)$  oscillates and departs from any eigenvalues; see Fig.~\ref{fig:enter-label3}(e) and its enlarged part in Fig.~\ref{fig:enter-label3}(f).
This is consistent with the analysis of projection probability.
In the intermediate nonlinear strength, there is only a left-to-right channel that can support topological pumping of the edge soliton. 
The energy bifurcation of the right-to-edge channel prevents topological pumping no matter how slow the modulation is.
The breakdown of adiabaticity in the presence of energy bifurcations is quite common.
Under periodic boundary condition, the breakdown of quantization in bulk Thouless pumping is also attributed to the presence of self-crossing in nonlinear topological bands~\cite{Tuloup_2023}.

Note that the nonlinear instantaneous eigenstates generally are not orthogonal to each other, and the sum of the projection probabilities will not be unity. 
In this case, even though the fidelity between the evolved state and a cetain nonlinear eigenstate is not zero, it does not indicate that the evolved state adiabatical follows this nonlinear eigenstate. 
Only when the sum of projection probabilities is $1$ and the projection probability onto the band of a certain eigenstate is close to $1$, we can safely claim that the evolved state adiabatically follows such nonlinear instantaneous eigenstates.

\begin{figure}[htpb]
    \centering
    \includegraphics[width=1.03\linewidth]{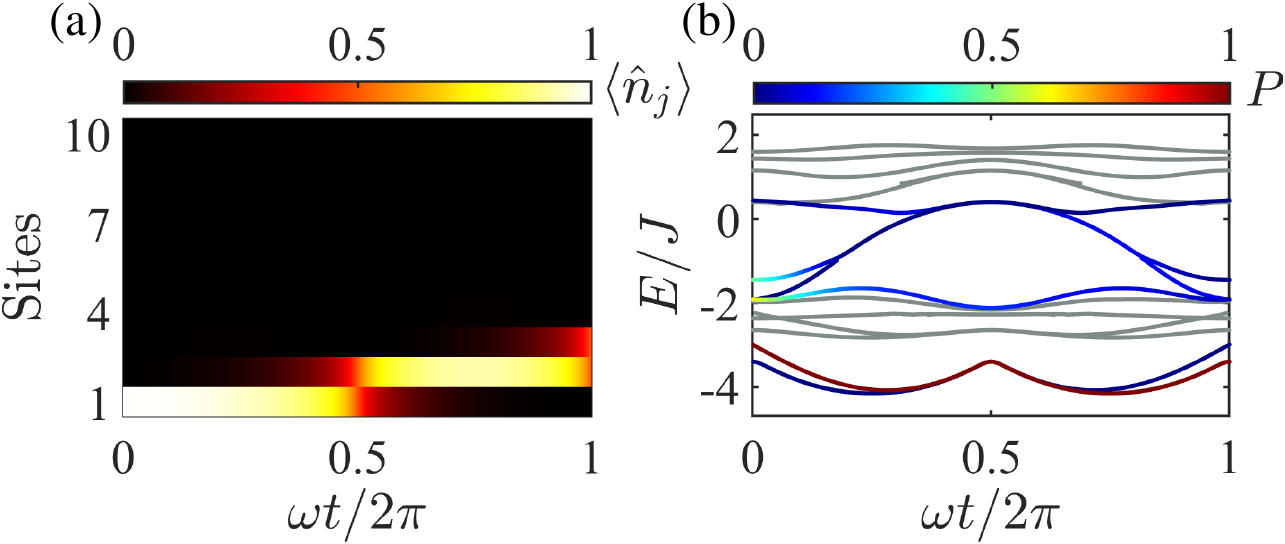}
    \caption{(a) Time evolution of density distribution in one pumping cycle with initial edge soliton completely localized at the left edge. (b) Nonlinear energy spectrum. The colors in (a) denote the density distribution, while the colors in (b) denote the probability of instantaneous nonlinear eigenstates occupying the dynamically evolved edge solitons. The other parameters are the same as those in Figs.~\ref{fig:enter-label2} except for $g=3$.}
    \label{fig:enter-label4}
\end{figure}

\subsection{Hybridized topological pumping} \label{hybridized}
 If the nonlinear strength is further enhanced ($g=3$), the pumping process from the right to the left edges also fails, similar to the case of $g=1.5$ in the previous subsection.
However, the topological pumping of the edge soliton from the left to the right edges is quite different from the case of $g=1.5$. 
This is because there exist extra more edge solitons for stronger interaction strength.
We can obtain two types of edge solitons (namely, thin and thick edge solitons) with different initial guesses. 
The thin edge soliton is completely located at a single lattice site, which is obtained using all particles at the edge site as an initial guess.
The thick edge soliton is less localized,  which is obtained using the edge soliton with $g=2.5$ at $t=0$ as the initial guess. 
The support thin edge soliton, the nonlinear strength should be larger than a critical value $g_{c,2}=2.4$, while the other parameters are the same as those in Fig.~\ref{fig:enter-label2}. 
We only focus on the topological pumping of thin edge solitons, because the pumping transport mechanism of thick edge solitons is identical to the one in Fig.~\ref{fig:enter-label3}.
To be specific, thick edge solitons can accomplish topological pumping from the left edge to the right edge, while reverse transport from the right edge to the left
edge is forbidden due to the existence of energy bifurcations in the upper right-to-left channel.

In one pumping cycle, an initial thin edge soliton can be transported from the left edge to the superposition of the second and third sites; as shown in Fig.~\ref{fig:enter-label4}(a).
This means that the picture of the left-to-right transport channel is broken. 
To understand what happens in this pumping process, we calculate the energy spectrum as a function of time with colors marking the projection probability of evolved states onto the instantaneous eigenstates; see Fig.~\ref{fig:enter-label4}(b).
We find that the thin edge soliton does not exist in the band gap between the two bulk bands.
As nonlinear strength increases, the thin edge soliton will pass through the lower bulk band and hybridize with one of the bulk states that separate from the other bulk states.
The two lowest modes correspond to edge and bulk solitons.
Surprisingly, the coupling between these two types of soliton is nonreciprocal in one pumping cycle. 
The thin edge soliton can be transferred to the bulk soliton in the second and third sites.
However, the bulk soilton in the second and third sites will be transferred to the other bulk soliton in the fourth and fifth sites, which obeys the rule of Thouless pumping of bulk states.

\begin{figure}[htpb]
    \centering
    \includegraphics[width=0.98\linewidth]{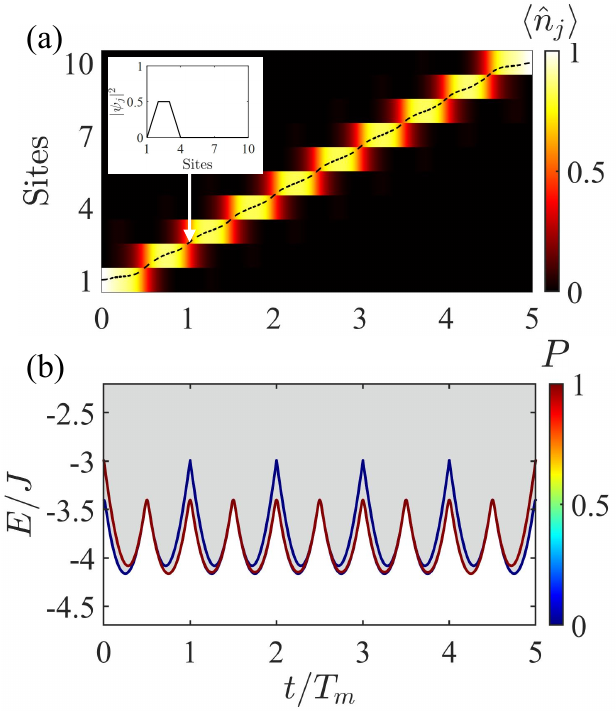}
    \caption{(a) Time evolution of density distribution in five pumping cycle with the same initial state as Fig.~\ref{fig:enter-label4}. The inset shows the probability distribution of the soliton after one cycle. The black dashed line denotes the mean position of the soliton. (b) Nonlinear energy spectrum, where the colors denote the probability of the evolved solitons occupying the instantaneous nonlinear eigenstates  and the grey background  denotes the region of bulk bands. The other parameters are the same as those in Fig.~\ref{fig:enter-label4}.}
    \label{fig:enter-label5}
\end{figure}

This property motivates us to perform hybridized topological pumping after multiple pumping cycles.
In Fig.~\ref{fig:enter-label5}(a), we show the evolution of the density distribution in five pumping cycles. 
After the thin edge soliton in the first site is transferred to the bulk soliton in the second and third sites; see its density distribution in the inset of Fig.~\ref{fig:enter-label5}(a).
In the next three pumping cycles this bulk soliton is transferred to the one in the eighth and ninth sites, and in the last pumping cycle the bulk soliton is transferred to the thin edge soliton in the last site.
In the whole pumping process, the evolved soliton is always localized, either at one site or at two sites.
We also calculate the center-of-mass position of the evolved soliton with $X(t)=\sum_{j}j|\psi_{j}(t)|^{2}$; see the black dashed line in Fig.~\ref{fig:enter-label5}(a). 
The displacement in the first and last pumping cycles is $1.5$, while the displacement in the middle three pumping cycles is $6$, which is governed by the Chern number $C$ of the lowest band in the linear case.
Similarly, we show the projection probability of the evolved states onto the two lowest modes; see Fig.~\ref{fig:enter-label5}(b).
The probability of the instantaneous state projected into the eigenstates is close to $P=1$, meaning that the wave function always evolves adiabatically along the nonlinear instantaneous eigenstates.
In particular, in the first and last pumping cycles, the energy of evolved soliton swaps from the thin edge soliton to the bulk soliton or vice versa, while in the middle pumping cycles, the energy of evolved soliton always stays in the lowest bulk mode.
Hybridized topological pumping is a new pattern of topological pumping that differs from both conventional topological pumping of edge states and Thouless pumping. 
The nonlinearity plays a crucial role in hybridizing the bulk and edge soliton.
For a general lattice with $2N$ sites, the thin edge soliton at the left end can be adiabatically transferred to the one at the right end after $N$ pumping cycles.
Because the soliton is always localized in the whole pumping process, such hybridized topological pumping can be applied for quantum state transfer. 

\begin{figure}[htpb]
    \centering
    \includegraphics[width=0.98\linewidth]{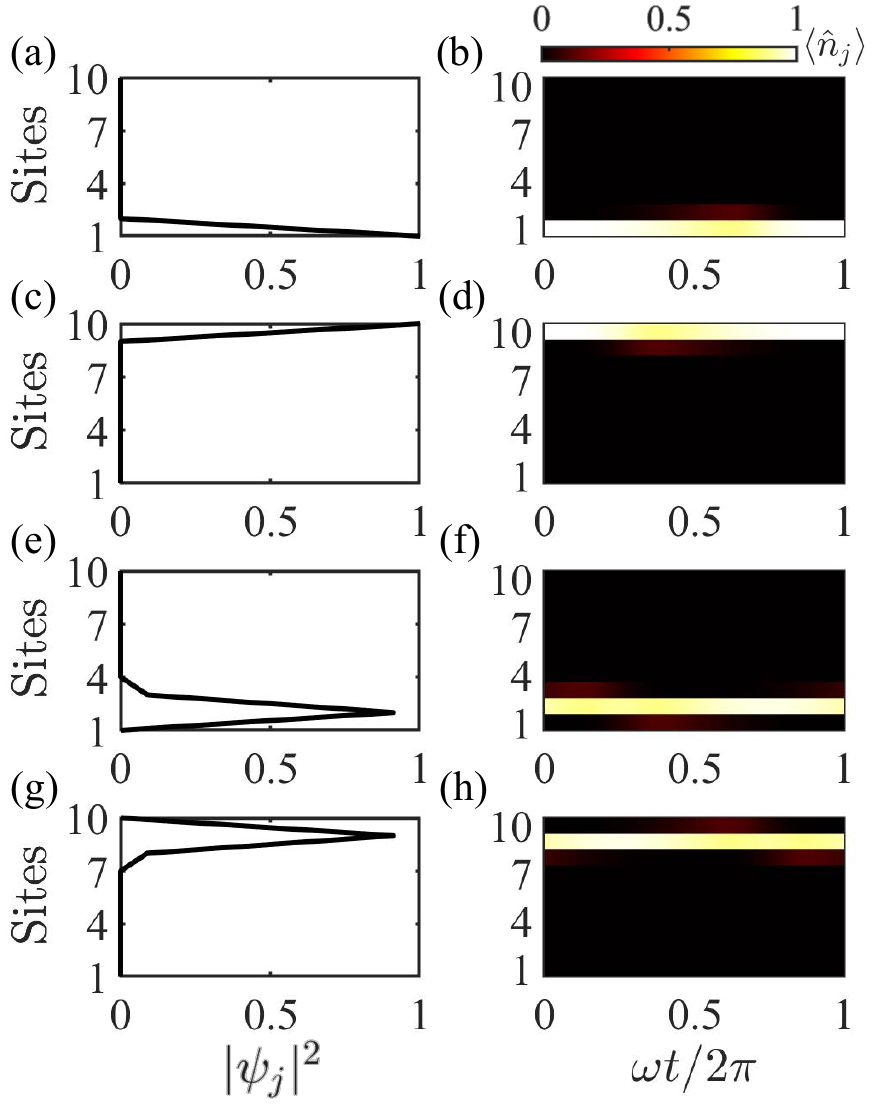}
    \caption{(a), (c): The soliton at $t=0$ is prepared in the left and right edges, respectively. 
    (b), (d): Time evolution of density distribution with initial states shown in (a), (c), respectively. 
    (e), (g): The sub-edge soliton at $t=0$ is prepared around the second and $9$th sites, respectively.
    (f), (h): Time evolution of sub-edge solitons density distribution  with initial sub-edge solitons shown in (e), (g), respectively.
    The parameters are the same as those in Fig.~\ref{fig:enter-label2} except for $g=7$.}
    \label{fig:enter-label6}
\end{figure}

\subsection{Self-trap} \label{selftrap}
\begin{figure}[!htp]
    \centering
    \includegraphics[width=\linewidth]{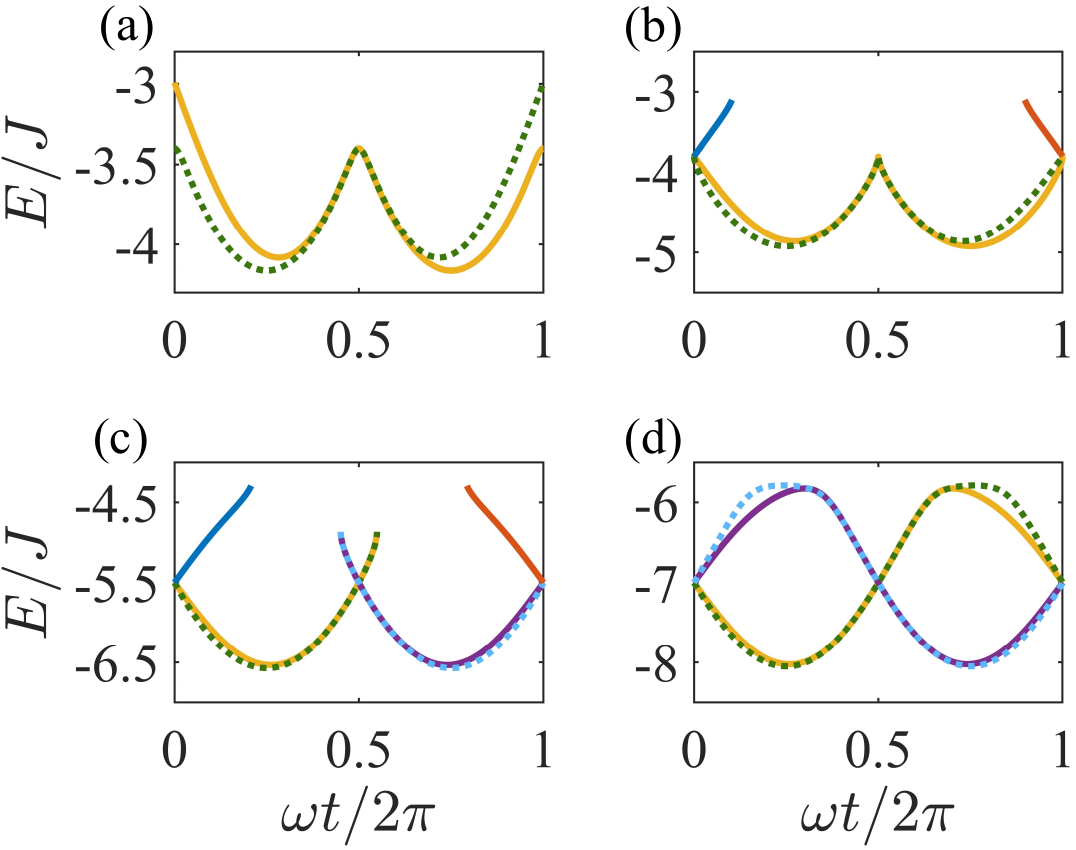}
    \caption{Nonlinear energy spectrum for different values of the nonlinear strength. The nonlinear strengthes are chosen are (a) $g=3$, (b) $g=3.8$, (c) $g=5.5$, (d) $g=7$, while the other parameters are the same as those in Figs.~\ref{fig:enter-label2}.}
    \label{fig:enter-label7}
\end{figure}

In Thouless pumping of bulk solitons, self-trapping has been observed under strong nonlinearity in experiments~\cite{jurgensen2021quantized,
jurgensen2023quantized}. 
However, there is no study of the fate of topological pumping of edge solitons under strong nonlinearity.
Using $g=7$ as an example, we find that there are four edge solitons, two edge solitons in the left and right sites, and two solitons in the sub-edge sites.
The emergence of the sub-edge solitons come from the interplay between interaction and topology.
This is quite different from the interaction-induced sub-edge states, where topological phase is trivial in the noninteracting case~\cite{Liu_2023}.
When initial states are prepared as the four edge solitons, these states will mostly stay at the initial position in one pumping cycle; see the time evolution of the density distributions in Figs.~\ref{fig:enter-label6}(b,d,f,h).
We find that all edge solitons are well localized and return to the initial states; see Figs.~\ref{fig:enter-label6}(a,c,e,g). 
We try to understand the transition from hybridized topological pumping to self-trap as nonlinearity increases.
Figs.~\ref{fig:enter-label7}(a)-(d) show nonlinear energy spectrum of the soliton in one pumping cycle for different values of the nonlinear strengths $g=3,~3.8,~5.5,~7$, respectively. 
The soliton states prepared at the edges will track the instantaneous localized stable soliton solutions. 
Thus, we can analyze pumping dynamics of edge solitons in terms of instantaneous nonlinear eigenstates. 
This strategy works well in previous sections when the nonlinear eigenvalues are continuous functions of time; see Fig.~\ref{fig:enter-label7}(a).
As nonlinear strength increases, two additional solitons appear in certain regions of the modulation phase.
At the critical value $g_{c,3}=3.8$, the bands originally marked with yellow solid line and green dotted line turn sharp at $t=T/2$, indicating that the second deviations of the energies with respect to time become discontinuous. 
At this point, the edge soliton can still evolve adiabatically along the instantaneous nonlinear eigenstates. 
However, crossing the threshold value and taking $g=5.5$ for example, the two continuous bands appear crossing structures where the adiabatic evolution paths are destroyed;  see Fig.~\ref{fig:enter-label7}(c).
To clearly show the crossing structures, we mark the two lower disconnected bands as yellow solid (green dotted) line and light blue dotted (purple solid) line. 
For the two new discontinuous bands, the eigenstate of the blue band at time $t=0$ corresponds to a thin edge soliton at the right edge. 
Due to symmetry, the red band at time $t=T$ corresponds to a thin edge soliton at the left edge.  
As nonlinear strength further increases, the blue band will approach the purple band and finally form a new continuous band. 
Similarly,  the yellow band will form a complete band with the right band.
Taking $g=7$ for example, the continuity of the energy bands throughout the cycle means the formation of new adiabatic evolution paths; see Fig.~\ref{fig:enter-label7}(d).
The yellow and purple bands at time $t=0$ correspond to thin edge solitons at the left edge and the right edge, respectively. 
These initial solitons will be trapped in the pumping process; see Figs.~\ref{fig:enter-label6}(b, d).
Similarly, the phenomena of self-trap in Figs.~\ref{fig:enter-label6}(f, h) are originated from the green and bright-blue bands.

\section{CONCLUSION}\label{CONCLUSION}
\par We have studied four types of pumping dynamics of edge soliton with different nonlinear strengths in a modulated superlattice.
We have proposed an appropriate iterative method for the selection of the initial guess, which helps to calculate the continuous nonlinear energy spectrum. 
With the help of the energy spectrum, we explain asymmetric topological pumping, where the left-to-right channel differs from the right-to-left channel, hybridized topological pumping which combines the transport of bulk soliton and edge solitons, and self-trap due to self-crossing structure.
In particular, in hybridized topological pumping, the edge soliton maintains localization, which may be used for quantum state transfer. 
Because the solitons can be transferred from edge to any bulk site, this is superior to the previous schemes that support state transfer between left and right boundaries, or between two bulk sites.

The interplay between nonlinearity and topology is an interesting topic that deserves further study. 
On the one hand, many novel non-interacting toplogical states such as Floquet topological insulators~\cite{PhysRevB.79.081406,PhysRevB.82.235114,
lindner2011floquet,rechtsman2013photonic,PhysRevB.109.224315}, high-order topological insulators~\cite{PhysRevLett.119.246402,
PhysRevLett.119.246401,Schindler2018-uw,serra2018observation,
peterson2018quantized,PhysRevLett.126.146802}, non-Abelian toplogical states~\cite{MOORE1991362,RevModPhys.80.1083,guo2021experimental}, have been theoretically predicted and experimentally realized.   
We should understand how interaction and nonlinearity affect these topological states.
On the other hand, there are some novel solitons in higher-dimensional atomic Bose-Einstein condensate, such as vortex solitons~\cite{PhysRevE.94.032202,Pang2018-aa,XU2024115043} and skyrmion solitons~\cite{perapechka2018soliton,LIU20133300}.
However, solitons in higher dimensions are always less stable than solitons in one dimension.
One may reversely study how topological properties can affect the behavior and stability of solitons in higher dimensions.

\begin{acknowledgments}
\par The authors acknowledge useful discussions with Boning Huang and Wenjie Liu. This work was supported by the National Key Research and Development Program of China (Grant No. 2022YFA1404104), the National Natural Science Foundation of China (Grants Nos. 12025509 and 12275365), the Key-Area Research and Development Program of Guangdong Province (Grant No. 2019B030330001), Guangdong Provincial Quantum Science Strategic Initiative (GDZX2305006,GDZX2304007), and the Natural Science Foundation of Guangdong(Grant No. 2023A1515012099).
\end{acknowledgments}

\begin{figure}[!htp]
    \centering
    \includegraphics[width=1\linewidth]{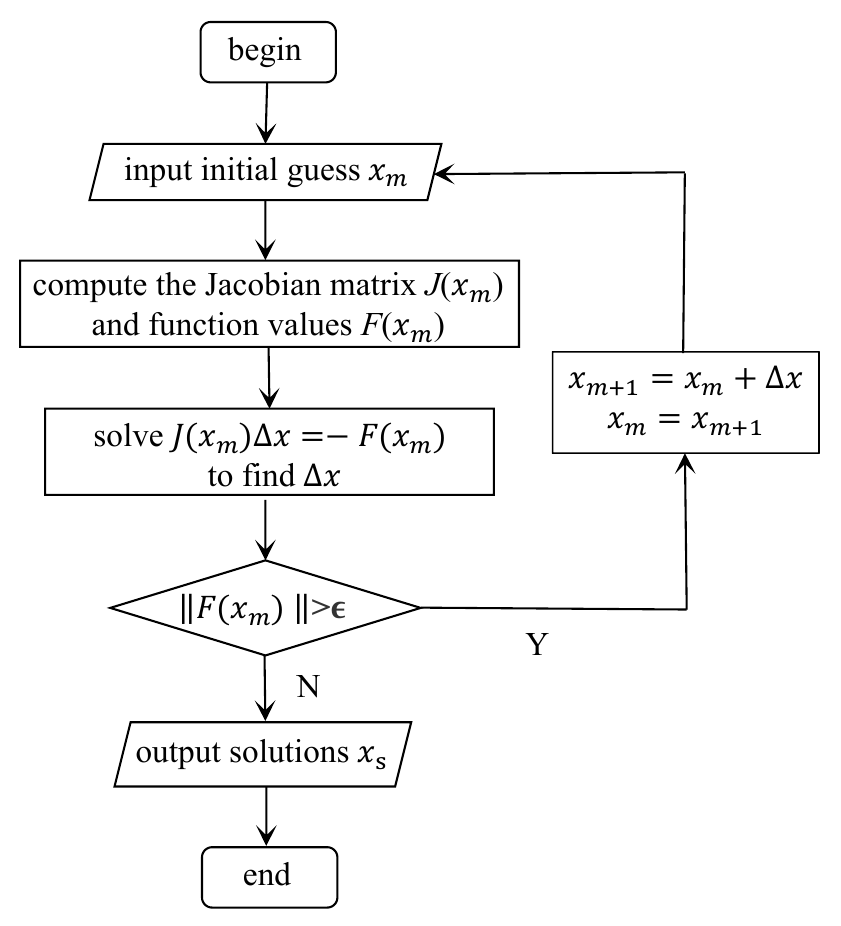}
    \caption{Numerical Newton-Jacobi iteration procedure for solving the nonlinear eigenvalues and eigenstates.}
    \label{fig:enter-label8}
\end{figure}

\appendix
\section{Numerical solution of instantaneous stationary states}\label{app1}

Newton-jacobi iterative method is a numerical method to solve nonlinear equations. 
This method combines the concepts of Newton's method and Jacobian matrix and aims to approach the root of the equation iteratively. 
Newton's method relies on Taylor expansion and linear approximation to quickly find an approximation of the root of an equation, while the Jacobian matrix provides a way to deal with functions of many variables, making the method widely used to solve multidimensional nonlinear problems.

We start with a numerical implementation of the one-variable Newton algorithm. We recall that the algorithm assumes the simple form $x_{m+1}=x_{m}-f(x_{m})/f^{\prime}(x_{m})$ to approximate the solution $x_{s}$ such that $f(x_{s})=0$. 
Here, $m$  denotes the iteration index and $f^{\prime}(x_{m})$ is the derivative of the function with respect to $x_{m}$. 
Next, we generalize the algorithm to the multi-variable case.
For a set of $M$ simultaneous equations $F=\{f_1,f_2,...,f_M\}$, there are $M$ unknown variables which can form a vector $x_m=\{x_m^{1}, x_m^{2}, x_m^{3},...,x_m^{M}\}$, where $x_m^{j}$ is the $j$th component of the vector.
The iteration algorithm can be generalized  to 
\begin{equation}
J(x_m) \cdot (x_{m+1}-x_{m})=-F(x_{m}),
\end{equation}
where $J$ is the Jacobian matrix, whose elements are given by $J_{i,j}=\partial f_j/\partial x_m^{j}$.
In Fig.~\ref{fig:enter-label8}, we show the key steps of the Newton-Jacobi iteration procedure to determine nonlinear eigenvalues and eigenstates. The four key steps are listed as follow.
\par 1. We choose initial guess of $x_{m}$ that is close to the root of the equations.
\par 2. We calculate the Jacobian matrix $J(x_m)$ and function $F(x_m)$ at the current guess  $x_{m}$.
\par 3. We solve $M$ linear equations $J(x_{m}) \Delta x=-F(x_{m})$ to find $\Delta x=x_{m+1}-x_m$.
\par 4. If $x_{m}$ tends to be the root of the equations, the norm of $F(x_{m})$ should be smaller than a threshold value $\epsilon$. If $F(x_{m})<\epsilon$, we stop the iteration and output $x_m+\Delta x$. Otherwise, we update the guess by $x_{m}=x_{m}+\Delta x$, go back to step $2$ and continue iteration.

Specifically, in our system with $N$ sites, the GP equation contains a set of $N$ simultaneous equations.
Taking into account the normalization condition $\sum_j |\psi_j|^2=1$, there are total $M=N+1$ equations.
Meanwhile, the vector with $M$ unknown variables is $x=[\psi_{1}, \cdots, \psi_{j}, \cdots, \psi_{N}, \lambda]^{T}$.
By setting the threshold value of $\epsilon$ as $10^{-10}$, we can precisely calculate the nonlinear eigenvalues and eigenstates with the Newton-Jacobi algorithm.

\nocite{*}


\begin{thebibliography}{88}%
\makeatletter
\providecommand \@ifxundefined [1]{%
 \@ifx{#1\undefined}
}%
\providecommand \@ifnum [1]{%
 \ifnum #1\expandafter \@firstoftwo
 \else \expandafter \@secondoftwo
 \fi
}%
\providecommand \@ifx [1]{%
 \ifx #1\expandafter \@firstoftwo
 \else \expandafter \@secondoftwo
 \fi
}%
\providecommand \natexlab [1]{#1}%
\providecommand \enquote  [1]{``#1''}%
\providecommand \bibnamefont  [1]{#1}%
\providecommand \bibfnamefont [1]{#1}%
\providecommand \citenamefont [1]{#1}%
\providecommand \href@noop [0]{\@secondoftwo}%
\providecommand \href [0]{\begingroup \@sanitize@url \@href}%
\providecommand \@href[1]{\@@startlink{#1}\@@href}%
\providecommand \@@href[1]{\endgroup#1\@@endlink}%
\providecommand \@sanitize@url [0]{\catcode `\\12\catcode `\$12\catcode
  `\&12\catcode `\#12\catcode `\^12\catcode `\_12\catcode `\%12\relax}%
\providecommand \@@startlink[1]{}%
\providecommand \@@endlink[0]{}%
\providecommand \url  [0]{\begingroup\@sanitize@url \@url }%
\providecommand \@url [1]{\endgroup\@href {#1}{\urlprefix }}%
\providecommand \urlprefix  [0]{URL }%
\providecommand \Eprint [0]{\href }%
\providecommand \doibase [0]{https://doi.org/}%
\providecommand \selectlanguage [0]{\@gobble}%
\providecommand \bibinfo  [0]{\@secondoftwo}%
\providecommand \bibfield  [0]{\@secondoftwo}%
\providecommand \translation [1]{[#1]}%
\providecommand \BibitemOpen [0]{}%
\providecommand \bibitemStop [0]{}%
\providecommand \bibitemNoStop [0]{.\EOS\space}%
\providecommand \EOS [0]{\spacefactor3000\relax}%
\providecommand \BibitemShut  [1]{\csname bibitem#1\endcsname}%
\let\auto@bib@innerbib\@empty
\bibitem [{\citenamefont {Citro}\ and\ \citenamefont
  {Aidelsburger}(2023)}]{citro2023thouless}%
  \BibitemOpen
  \bibfield  {author} {\bibinfo {author} {\bibfnamefont {R.}~\bibnamefont
  {Citro}}\ and\ \bibinfo {author} {\bibfnamefont {M.}~\bibnamefont
  {Aidelsburger}},\ }\bibfield  {title} {\bibinfo {title} {{Thouless} pumping
  and topology},\ }\href {https://www.nature.com/articles/s42254-022-00545-0}
  {\bibfield  {journal} {\bibinfo  {journal} {Nat. Rev. Phys.}\ }\textbf
  {\bibinfo {volume} {5}},\ \bibinfo {pages} {87} (\bibinfo {year}
  {2023})}\BibitemShut {NoStop}%
\bibitem [{\citenamefont {Thouless}(1983)}]{PhysRevB.27.6083}%
  \BibitemOpen
  \bibfield  {author} {\bibinfo {author} {\bibfnamefont {D.~J.}\ \bibnamefont
  {Thouless}},\ }\bibfield  {title} {\bibinfo {title} {Quantization of particle
  transport},\ }\href {https://doi.org/10.1103/PhysRevB.27.6083} {\bibfield
  {journal} {\bibinfo  {journal} {Phys. Rev. B}\ }\textbf {\bibinfo {volume}
  {27}},\ \bibinfo {pages} {6083} (\bibinfo {year} {1983})}\BibitemShut
  {NoStop}%
\bibitem [{\citenamefont {Kraus}\ \emph {et~al.}(2012)\citenamefont {Kraus},
  \citenamefont {Lahini}, \citenamefont {Ringel}, \citenamefont {Verbin},\ and\
  \citenamefont {Zilberberg}}]{PhysRevLett.109.106402}%
  \BibitemOpen
  \bibfield  {author} {\bibinfo {author} {\bibfnamefont {Y.~E.}\ \bibnamefont
  {Kraus}}, \bibinfo {author} {\bibfnamefont {Y.}~\bibnamefont {Lahini}},
  \bibinfo {author} {\bibfnamefont {Z.}~\bibnamefont {Ringel}}, \bibinfo
  {author} {\bibfnamefont {M.}~\bibnamefont {Verbin}},\ and\ \bibinfo {author}
  {\bibfnamefont {O.}~\bibnamefont {Zilberberg}},\ }\bibfield  {title}
  {\bibinfo {title} {Topological states and adiabatic pumping in
  quasicrystals},\ }\href {https://doi.org/10.1103/PhysRevLett.109.106402}
  {\bibfield  {journal} {\bibinfo  {journal} {Phys. Rev. Lett.}\ }\textbf
  {\bibinfo {volume} {109}},\ \bibinfo {pages} {106402} (\bibinfo {year}
  {2012})}\BibitemShut {NoStop}%
\bibitem [{\citenamefont {Cheng}\ \emph {et~al.}(2022)\citenamefont {Cheng},
  \citenamefont {Wang}, \citenamefont {Ke}, \citenamefont {Chen}, \citenamefont
  {Yu}, \citenamefont {Kivshar}, \citenamefont {Lee},\ and\ \citenamefont
  {Pan}}]{cheng2022asymmetric}%
  \BibitemOpen
  \bibfield  {author} {\bibinfo {author} {\bibfnamefont {Q.}~\bibnamefont
  {Cheng}}, \bibinfo {author} {\bibfnamefont {H.}~\bibnamefont {Wang}},
  \bibinfo {author} {\bibfnamefont {Y.}~\bibnamefont {Ke}}, \bibinfo {author}
  {\bibfnamefont {T.}~\bibnamefont {Chen}}, \bibinfo {author} {\bibfnamefont
  {Y.}~\bibnamefont {Yu}}, \bibinfo {author} {\bibfnamefont {Y.~S.}\
  \bibnamefont {Kivshar}}, \bibinfo {author} {\bibfnamefont {C.}~\bibnamefont
  {Lee}},\ and\ \bibinfo {author} {\bibfnamefont {Y.}~\bibnamefont {Pan}},\
  }\bibfield  {title} {\bibinfo {title} {Asymmetric topological pumping in
  nonparaxial photonics},\ }\href
  {https://www.nature.com/articles/s41467-021-27773-9} {\bibfield  {journal}
  {\bibinfo  {journal} {Nat. Commun.}\ }\textbf {\bibinfo {volume} {13}},\
  \bibinfo {pages} {249} (\bibinfo {year} {2022})}\BibitemShut {NoStop}%
\bibitem [{\citenamefont {King-Smith}\ and\ \citenamefont
  {Vanderbilt}(1993)}]{PhysRevB.47.1651}%
  \BibitemOpen
  \bibfield  {author} {\bibinfo {author} {\bibfnamefont {R.~D.}\ \bibnamefont
  {King-Smith}}\ and\ \bibinfo {author} {\bibfnamefont {D.}~\bibnamefont
  {Vanderbilt}},\ }\bibfield  {title} {\bibinfo {title} {Theory of polarization
  of crystalline solids},\ }\href {https://doi.org/10.1103/PhysRevB.47.1651}
  {\bibfield  {journal} {\bibinfo  {journal} {Phys. Rev. B}\ }\textbf {\bibinfo
  {volume} {47}},\ \bibinfo {pages} {1651} (\bibinfo {year}
  {1993})}\BibitemShut {NoStop}%
\bibitem [{\citenamefont {Xiao}\ \emph {et~al.}(2010)\citenamefont {Xiao},
  \citenamefont {Chang},\ and\ \citenamefont {Niu}}]{RevModPhys.82.1959}%
  \BibitemOpen
  \bibfield  {author} {\bibinfo {author} {\bibfnamefont {D.}~\bibnamefont
  {Xiao}}, \bibinfo {author} {\bibfnamefont {M.-C.}\ \bibnamefont {Chang}},\
  and\ \bibinfo {author} {\bibfnamefont {Q.}~\bibnamefont {Niu}},\ }\bibfield
  {title} {\bibinfo {title} {{Berry} phase effects on electronic properties},\
  }\href {https://doi.org/10.1103/RevModPhys.82.1959} {\bibfield  {journal}
  {\bibinfo  {journal} {Rev. Mod. Phys.}\ }\textbf {\bibinfo {volume} {82}},\
  \bibinfo {pages} {1959} (\bibinfo {year} {2010})}\BibitemShut {NoStop}%
\bibitem [{\citenamefont {Lohse}\ \emph {et~al.}(2016)\citenamefont {Lohse},
  \citenamefont {Schweizer}, \citenamefont {Zilberberg}, \citenamefont
  {Aidelsburger},\ and\ \citenamefont {Bloch}}]{lohse2016thouless}%
  \BibitemOpen
  \bibfield  {author} {\bibinfo {author} {\bibfnamefont {M.}~\bibnamefont
  {Lohse}}, \bibinfo {author} {\bibfnamefont {C.}~\bibnamefont {Schweizer}},
  \bibinfo {author} {\bibfnamefont {O.}~\bibnamefont {Zilberberg}}, \bibinfo
  {author} {\bibfnamefont {M.}~\bibnamefont {Aidelsburger}},\ and\ \bibinfo
  {author} {\bibfnamefont {I.}~\bibnamefont {Bloch}},\ }\bibfield  {title}
  {\bibinfo {title} {A {Thouless} quantum pump with ultracold bosonic atoms in
  an optical superlattice},\ }\href {https://www.nature.com/articles/nphys3584}
  {\bibfield  {journal} {\bibinfo  {journal} {Nat. Phys.}\ }\textbf {\bibinfo
  {volume} {12}},\ \bibinfo {pages} {350} (\bibinfo {year} {2016})}\BibitemShut
  {NoStop}%
\bibitem [{\citenamefont {Nakajima}\ \emph {et~al.}(2016)\citenamefont
  {Nakajima}, \citenamefont {Tomita}, \citenamefont {Taie}, \citenamefont
  {Ichinose}, \citenamefont {Ozawa}, \citenamefont {Wang}, \citenamefont
  {Troyer},\ and\ \citenamefont {Takahashi}}]{nakajima2016topological}%
  \BibitemOpen
  \bibfield  {author} {\bibinfo {author} {\bibfnamefont {S.}~\bibnamefont
  {Nakajima}}, \bibinfo {author} {\bibfnamefont {T.}~\bibnamefont {Tomita}},
  \bibinfo {author} {\bibfnamefont {S.}~\bibnamefont {Taie}}, \bibinfo {author}
  {\bibfnamefont {T.}~\bibnamefont {Ichinose}}, \bibinfo {author}
  {\bibfnamefont {H.}~\bibnamefont {Ozawa}}, \bibinfo {author} {\bibfnamefont
  {L.}~\bibnamefont {Wang}}, \bibinfo {author} {\bibfnamefont {M.}~\bibnamefont
  {Troyer}},\ and\ \bibinfo {author} {\bibfnamefont {Y.}~\bibnamefont
  {Takahashi}},\ }\bibfield  {title} {\bibinfo {title} {Topological {Thouless}
  pumping of ultracold fermions},\ }\href
  {https://www.nature.com/articles/nphys3622} {\bibfield  {journal} {\bibinfo
  {journal} {Nat. Phys.}\ }\textbf {\bibinfo {volume} {12}},\ \bibinfo {pages}
  {296} (\bibinfo {year} {2016})}\BibitemShut {NoStop}%
\bibitem [{\citenamefont {Wang}\ \emph {et~al.}(2013)\citenamefont {Wang},
  \citenamefont {Troyer},\ and\ \citenamefont {Dai}}]{PhysRevLett.111.026802}%
  \BibitemOpen
  \bibfield  {author} {\bibinfo {author} {\bibfnamefont {L.}~\bibnamefont
  {Wang}}, \bibinfo {author} {\bibfnamefont {M.}~\bibnamefont {Troyer}},\ and\
  \bibinfo {author} {\bibfnamefont {X.}~\bibnamefont {Dai}},\ }\bibfield
  {title} {\bibinfo {title} {Topological charge pumping in a one-dimensional
  optical lattice},\ }\href {https://doi.org/10.1103/PhysRevLett.111.026802}
  {\bibfield  {journal} {\bibinfo  {journal} {Phys. Rev. Lett.}\ }\textbf
  {\bibinfo {volume} {111}},\ \bibinfo {pages} {026802} (\bibinfo {year}
  {2013})}\BibitemShut {NoStop}%
\bibitem [{\citenamefont {Verbin}\ \emph {et~al.}(2015)\citenamefont {Verbin},
  \citenamefont {Zilberberg}, \citenamefont {Lahini}, \citenamefont {Kraus},\
  and\ \citenamefont {Silberberg}}]{PhysRevB.91.064201}%
  \BibitemOpen
  \bibfield  {author} {\bibinfo {author} {\bibfnamefont {M.}~\bibnamefont
  {Verbin}}, \bibinfo {author} {\bibfnamefont {O.}~\bibnamefont {Zilberberg}},
  \bibinfo {author} {\bibfnamefont {Y.}~\bibnamefont {Lahini}}, \bibinfo
  {author} {\bibfnamefont {Y.~E.}\ \bibnamefont {Kraus}},\ and\ \bibinfo
  {author} {\bibfnamefont {Y.}~\bibnamefont {Silberberg}},\ }\bibfield  {title}
  {\bibinfo {title} {Topological pumping over a photonic {Fibonacci}
  quasicrystal},\ }\href {https://doi.org/10.1103/PhysRevB.91.064201}
  {\bibfield  {journal} {\bibinfo  {journal} {Phys. Rev. B}\ }\textbf {\bibinfo
  {volume} {91}},\ \bibinfo {pages} {064201} (\bibinfo {year}
  {2015})}\BibitemShut {NoStop}%
\bibitem [{\citenamefont {Ke}\ \emph {et~al.}(2016)\citenamefont {Ke},
  \citenamefont {Qin}, \citenamefont {Mei}, \citenamefont {Zhong},
  \citenamefont {Kivshar},\ and\ \citenamefont {Lee}}]{KeLPR2016}%
  \BibitemOpen
  \bibfield  {author} {\bibinfo {author} {\bibfnamefont {Y.}~\bibnamefont
  {Ke}}, \bibinfo {author} {\bibfnamefont {X.}~\bibnamefont {Qin}}, \bibinfo
  {author} {\bibfnamefont {F.}~\bibnamefont {Mei}}, \bibinfo {author}
  {\bibfnamefont {H.}~\bibnamefont {Zhong}}, \bibinfo {author} {\bibfnamefont
  {Y.~S.}\ \bibnamefont {Kivshar}},\ and\ \bibinfo {author} {\bibfnamefont
  {C.}~\bibnamefont {Lee}},\ }\bibfield  {title} {\bibinfo {title} {Topological
  phase transitions and {Thouless} pumping of light in photonic waveguide
  arrays},\ }\href {https://doi.org/https://doi.org/10.1002/lpor.201600119}
  {\bibfield  {journal} {\bibinfo  {journal} {Laser Photonics Rev.}\ }\textbf
  {\bibinfo {volume} {10}},\ \bibinfo {pages} {995} (\bibinfo {year}
  {2016})}\BibitemShut {NoStop}%
\bibitem [{\citenamefont {Cerjan}\ \emph {et~al.}(2020)\citenamefont {Cerjan},
  \citenamefont {Wang}, \citenamefont {Huang}, \citenamefont {Chen},\ and\
  \citenamefont {Rechtsman}}]{cerjan2020thouless}%
  \BibitemOpen
  \bibfield  {author} {\bibinfo {author} {\bibfnamefont {A.}~\bibnamefont
  {Cerjan}}, \bibinfo {author} {\bibfnamefont {M.}~\bibnamefont {Wang}},
  \bibinfo {author} {\bibfnamefont {S.}~\bibnamefont {Huang}}, \bibinfo
  {author} {\bibfnamefont {K.~P.}\ \bibnamefont {Chen}},\ and\ \bibinfo
  {author} {\bibfnamefont {M.~C.}\ \bibnamefont {Rechtsman}},\ }\bibfield
  {title} {\bibinfo {title} {{Thouless} pumping in disordered photonic
  systems},\ }\href {https://www.nature.com/articles/s41377-020-00408-2}
  {\bibfield  {journal} {\bibinfo  {journal} {Light Sci. Appl.}\ }\textbf
  {\bibinfo {volume} {9}},\ \bibinfo {pages} {178} (\bibinfo {year}
  {2020})}\BibitemShut {NoStop}%
\bibitem [{\citenamefont {Sun}\ \emph {et~al.}(2022)\citenamefont {Sun},
  \citenamefont {Zhang}, \citenamefont {Yu}, \citenamefont {Tian},
  \citenamefont {Chen},\ and\ \citenamefont {Sun}}]{sun2022non}%
  \BibitemOpen
  \bibfield  {author} {\bibinfo {author} {\bibfnamefont {Y.-K.}\ \bibnamefont
  {Sun}}, \bibinfo {author} {\bibfnamefont {X.-L.}\ \bibnamefont {Zhang}},
  \bibinfo {author} {\bibfnamefont {F.}~\bibnamefont {Yu}}, \bibinfo {author}
  {\bibfnamefont {Z.-N.}\ \bibnamefont {Tian}}, \bibinfo {author}
  {\bibfnamefont {Q.-D.}\ \bibnamefont {Chen}},\ and\ \bibinfo {author}
  {\bibfnamefont {H.-B.}\ \bibnamefont {Sun}},\ }\bibfield  {title} {\bibinfo
  {title} {Non-abelian {Thouless} pumping in photonic waveguides},\ }\href
  {https://www.nature.com/articles/s41567-022-01669-x} {\bibfield  {journal}
  {\bibinfo  {journal} {Nat. Phys.}\ }\textbf {\bibinfo {volume} {18}},\
  \bibinfo {pages} {1080} (\bibinfo {year} {2022})}\BibitemShut {NoStop}%
\bibitem [{\citenamefont {Ma}\ \emph {et~al.}(2018)\citenamefont {Ma},
  \citenamefont {Zhou}, \citenamefont {Zhang}, \citenamefont {Li},
  \citenamefont {Cheng}, \citenamefont {Geng}, \citenamefont {Rong},
  \citenamefont {Shi}, \citenamefont {Gong},\ and\ \citenamefont
  {Du}}]{PhysRevLett.120.120501}%
  \BibitemOpen
  \bibfield  {author} {\bibinfo {author} {\bibfnamefont {W.}~\bibnamefont
  {Ma}}, \bibinfo {author} {\bibfnamefont {L.}~\bibnamefont {Zhou}}, \bibinfo
  {author} {\bibfnamefont {Q.}~\bibnamefont {Zhang}}, \bibinfo {author}
  {\bibfnamefont {M.}~\bibnamefont {Li}}, \bibinfo {author} {\bibfnamefont
  {C.}~\bibnamefont {Cheng}}, \bibinfo {author} {\bibfnamefont
  {J.}~\bibnamefont {Geng}}, \bibinfo {author} {\bibfnamefont {X.}~\bibnamefont
  {Rong}}, \bibinfo {author} {\bibfnamefont {F.}~\bibnamefont {Shi}}, \bibinfo
  {author} {\bibfnamefont {J.}~\bibnamefont {Gong}},\ and\ \bibinfo {author}
  {\bibfnamefont {J.}~\bibnamefont {Du}},\ }\bibfield  {title} {\bibinfo
  {title} {Experimental observation of a generalized {Thouless} pump with a
  single spin},\ }\href {https://doi.org/10.1103/PhysRevLett.120.120501}
  {\bibfield  {journal} {\bibinfo  {journal} {Phys. Rev. Lett.}\ }\textbf
  {\bibinfo {volume} {120}},\ \bibinfo {pages} {120501} (\bibinfo {year}
  {2018})}\BibitemShut {NoStop}%
\bibitem [{\citenamefont {Alicea}\ \emph {et~al.}(2011)\citenamefont {Alicea},
  \citenamefont {Oreg}, \citenamefont {Refael}, \citenamefont {Von~Oppen},\
  and\ \citenamefont {Fisher}}]{alicea2011non}%
  \BibitemOpen
  \bibfield  {author} {\bibinfo {author} {\bibfnamefont {J.}~\bibnamefont
  {Alicea}}, \bibinfo {author} {\bibfnamefont {Y.}~\bibnamefont {Oreg}},
  \bibinfo {author} {\bibfnamefont {G.}~\bibnamefont {Refael}}, \bibinfo
  {author} {\bibfnamefont {F.}~\bibnamefont {Von~Oppen}},\ and\ \bibinfo
  {author} {\bibfnamefont {M.~P.}\ \bibnamefont {Fisher}},\ }\bibfield  {title}
  {\bibinfo {title} {{Non-Abelian} statistics and topological quantum
  information processing in {1D} wire networks},\ }\href
  {https://www.nature.com/articles/nphys1915} {\bibfield  {journal} {\bibinfo
  {journal} {Nat. Phys.}\ }\textbf {\bibinfo {volume} {7}},\ \bibinfo {pages}
  {412} (\bibinfo {year} {2011})}\BibitemShut {NoStop}%
\bibitem [{\citenamefont {Nayak}\ \emph
  {et~al.}(2008{\natexlab{a}})\citenamefont {Nayak}, \citenamefont {Simon},
  \citenamefont {Stern}, \citenamefont {Freedman},\ and\ \citenamefont
  {Das~Sarma}}]{nayak2008non}%
  \BibitemOpen
  \bibfield  {author} {\bibinfo {author} {\bibfnamefont {C.}~\bibnamefont
  {Nayak}}, \bibinfo {author} {\bibfnamefont {S.~H.}\ \bibnamefont {Simon}},
  \bibinfo {author} {\bibfnamefont {A.}~\bibnamefont {Stern}}, \bibinfo
  {author} {\bibfnamefont {M.}~\bibnamefont {Freedman}},\ and\ \bibinfo
  {author} {\bibfnamefont {S.}~\bibnamefont {Das~Sarma}},\ }\bibfield  {title}
  {\bibinfo {title} {{Non-Abelian} anyons and topological quantum
  computation},\ }\href
  {https://journals.aps.org/rmp/abstract/10.1103/RevModPhys.80.1083} {\bibfield
   {journal} {\bibinfo  {journal} {Rev. Mod. Phys.}\ }\textbf {\bibinfo
  {volume} {80}},\ \bibinfo {pages} {1083} (\bibinfo {year}
  {2008}{\natexlab{a}})}\BibitemShut {NoStop}%
\bibitem [{\citenamefont {Lang}\ and\ \citenamefont
  {B{\"u}chler}(2017)}]{lang2017topological}%
  \BibitemOpen
  \bibfield  {author} {\bibinfo {author} {\bibfnamefont {N.}~\bibnamefont
  {Lang}}\ and\ \bibinfo {author} {\bibfnamefont {H.~P.}\ \bibnamefont
  {B{\"u}chler}},\ }\bibfield  {title} {\bibinfo {title} {Topological networks
  for quantum communication between distant qubits},\ }\href
  {https://www.nature.com/articles/s41534-017-0047-x#citeas} {\bibfield
  {journal} {\bibinfo  {journal} {npj Quantum Inf.}\ }\textbf {\bibinfo
  {volume} {3}},\ \bibinfo {pages} {47} (\bibinfo {year} {2017})}\BibitemShut
  {NoStop}%
\bibitem [{\citenamefont {Bello}\ \emph {et~al.}(2016)\citenamefont {Bello},
  \citenamefont {Creffield},\ and\ \citenamefont {Platero}}]{bello2016long}%
  \BibitemOpen
  \bibfield  {author} {\bibinfo {author} {\bibfnamefont {M.}~\bibnamefont
  {Bello}}, \bibinfo {author} {\bibfnamefont {C.~E.}\ \bibnamefont
  {Creffield}},\ and\ \bibinfo {author} {\bibfnamefont {G.}~\bibnamefont
  {Platero}},\ }\bibfield  {title} {\bibinfo {title} {Long-range doublon
  transfer in a dimer chain induced by topology and ac fields},\ }\href
  {https://www.nature.com/articles/srep22562} {\bibfield  {journal} {\bibinfo
  {journal} {Scientific Reports}\ }\textbf {\bibinfo {volume} {6}},\ \bibinfo
  {pages} {22562} (\bibinfo {year} {2016})}\BibitemShut {NoStop}%
\bibitem [{\citenamefont {Dlaska}\ \emph {et~al.}(2017)\citenamefont {Dlaska},
  \citenamefont {Vermersch},\ and\ \citenamefont {Zoller}}]{dlaska2017robust}%
  \BibitemOpen
  \bibfield  {author} {\bibinfo {author} {\bibfnamefont {C.}~\bibnamefont
  {Dlaska}}, \bibinfo {author} {\bibfnamefont {B.}~\bibnamefont {Vermersch}},\
  and\ \bibinfo {author} {\bibfnamefont {P.}~\bibnamefont {Zoller}},\
  }\bibfield  {title} {\bibinfo {title} {Robust quantum state transfer via
  topologically protected edge channels in dipolar arrays},\ }\href
  {https://iopscience.iop.org/article/10.1088/2058-9565/2/1/015001} {\bibfield
  {journal} {\bibinfo  {journal} {Quantum Sci. Technol.}\ }\textbf {\bibinfo
  {volume} {2}},\ \bibinfo {pages} {015001} (\bibinfo {year}
  {2017})}\BibitemShut {NoStop}%
\bibitem [{\citenamefont {Mei}\ \emph {et~al.}(2018)\citenamefont {Mei},
  \citenamefont {Chen}, \citenamefont {Tian}, \citenamefont {Zhu},\ and\
  \citenamefont {Jia}}]{PhysRevA.98.012331}%
  \BibitemOpen
  \bibfield  {author} {\bibinfo {author} {\bibfnamefont {F.}~\bibnamefont
  {Mei}}, \bibinfo {author} {\bibfnamefont {G.}~\bibnamefont {Chen}}, \bibinfo
  {author} {\bibfnamefont {L.}~\bibnamefont {Tian}}, \bibinfo {author}
  {\bibfnamefont {S.-L.}\ \bibnamefont {Zhu}},\ and\ \bibinfo {author}
  {\bibfnamefont {S.}~\bibnamefont {Jia}},\ }\bibfield  {title} {\bibinfo
  {title} {Robust quantum state transfer via topological edge states in
  superconducting qubit chains},\ }\href
  {https://doi.org/10.1103/PhysRevA.98.012331} {\bibfield  {journal} {\bibinfo
  {journal} {Phys. Rev. A}\ }\textbf {\bibinfo {volume} {98}},\ \bibinfo
  {pages} {012331} (\bibinfo {year} {2018})}\BibitemShut {NoStop}%
\bibitem [{\citenamefont {Longhi}\ \emph {et~al.}(2019)\citenamefont {Longhi},
  \citenamefont {Giorgi},\ and\ \citenamefont {Zambrini}}]{longhi2019landau}%
  \BibitemOpen
  \bibfield  {author} {\bibinfo {author} {\bibfnamefont {S.}~\bibnamefont
  {Longhi}}, \bibinfo {author} {\bibfnamefont {G.~L.}\ \bibnamefont {Giorgi}},\
  and\ \bibinfo {author} {\bibfnamefont {R.}~\bibnamefont {Zambrini}},\
  }\bibfield  {title} {\bibinfo {title} {{Landau-Zener} topological quantum
  state transfer},\ }\href
  {https://onlinelibrary.wiley.com/doi/10.1002/qute.201800090} {\bibfield
  {journal} {\bibinfo  {journal} {Adv. Quantum Technol.}\ }\textbf {\bibinfo
  {volume} {2}},\ \bibinfo {pages} {1800090} (\bibinfo {year}
  {2019})}\BibitemShut {NoStop}%
\bibitem [{\citenamefont {Longhi}(2019)}]{PhysRevB.99.155150}%
  \BibitemOpen
  \bibfield  {author} {\bibinfo {author} {\bibfnamefont {S.}~\bibnamefont
  {Longhi}},\ }\bibfield  {title} {\bibinfo {title} {Topological pumping of
  edge states via adiabatic passage},\ }\href
  {https://doi.org/10.1103/PhysRevB.99.155150} {\bibfield  {journal} {\bibinfo
  {journal} {Phys. Rev. B}\ }\textbf {\bibinfo {volume} {99}},\ \bibinfo
  {pages} {155150} (\bibinfo {year} {2019})}\BibitemShut {NoStop}%
\bibitem [{\citenamefont {Qi}\ \emph {et~al.}(2020)\citenamefont {Qi},
  \citenamefont {Wang}, \citenamefont {Liu}, \citenamefont {Zhang},\ and\
  \citenamefont {Wang}}]{Qi:20}%
  \BibitemOpen
  \bibfield  {author} {\bibinfo {author} {\bibfnamefont {L.}~\bibnamefont
  {Qi}}, \bibinfo {author} {\bibfnamefont {G.-L.}\ \bibnamefont {Wang}},
  \bibinfo {author} {\bibfnamefont {S.}~\bibnamefont {Liu}}, \bibinfo {author}
  {\bibfnamefont {S.}~\bibnamefont {Zhang}},\ and\ \bibinfo {author}
  {\bibfnamefont {H.-F.}\ \bibnamefont {Wang}},\ }\bibfield  {title} {\bibinfo
  {title} {Controllable photonic and phononic topological state transfers in a
  small optomechanical lattice},\ }\href {https://doi.org/10.1364/OL.388835}
  {\bibfield  {journal} {\bibinfo  {journal} {Opt. Lett.}\ }\textbf {\bibinfo
  {volume} {45}},\ \bibinfo {pages} {2018} (\bibinfo {year}
  {2020})}\BibitemShut {NoStop}%
\bibitem [{\citenamefont {D'Angelis}\ \emph {et~al.}(2020)\citenamefont
  {D'Angelis}, \citenamefont {Pinheiro}, \citenamefont {Gu\'ery-Odelin},
  \citenamefont {Longhi},\ and\ \citenamefont
  {Impens}}]{PhysRevResearch.2.033475}%
  \BibitemOpen
  \bibfield  {author} {\bibinfo {author} {\bibfnamefont {F.~M.}\ \bibnamefont
  {D'Angelis}}, \bibinfo {author} {\bibfnamefont {F.~A.}\ \bibnamefont
  {Pinheiro}}, \bibinfo {author} {\bibfnamefont {D.}~\bibnamefont
  {Gu\'ery-Odelin}}, \bibinfo {author} {\bibfnamefont {S.}~\bibnamefont
  {Longhi}},\ and\ \bibinfo {author} {\bibfnamefont {F.~m.~c.}\ \bibnamefont
  {Impens}},\ }\bibfield  {title} {\bibinfo {title} {Fast and robust quantum
  state transfer in a topological {Su-Schrieffer-Heeger} chain with
  next-to-nearest-neighbor interactions},\ }\href
  {https://doi.org/10.1103/PhysRevResearch.2.033475} {\bibfield  {journal}
  {\bibinfo  {journal} {Phys. Rev. Res.}\ }\textbf {\bibinfo {volume} {2}},\
  \bibinfo {pages} {033475} (\bibinfo {year} {2020})}\BibitemShut {NoStop}%
\bibitem [{\citenamefont {Palaiodimopoulos}\ \emph {et~al.}(2021)\citenamefont
  {Palaiodimopoulos}, \citenamefont {Brouzos}, \citenamefont {Diakonos},\ and\
  \citenamefont {Theocharis}}]{PhysRevA.103.052409}%
  \BibitemOpen
  \bibfield  {author} {\bibinfo {author} {\bibfnamefont {N.~E.}\ \bibnamefont
  {Palaiodimopoulos}}, \bibinfo {author} {\bibfnamefont {I.}~\bibnamefont
  {Brouzos}}, \bibinfo {author} {\bibfnamefont {F.~K.}\ \bibnamefont
  {Diakonos}},\ and\ \bibinfo {author} {\bibfnamefont {G.}~\bibnamefont
  {Theocharis}},\ }\bibfield  {title} {\bibinfo {title} {Fast and robust
  quantum state transfer via a topological chain},\ }\href
  {https://doi.org/10.1103/PhysRevA.103.052409} {\bibfield  {journal} {\bibinfo
   {journal} {Phys. Rev. A}\ }\textbf {\bibinfo {volume} {103}},\ \bibinfo
  {pages} {052409} (\bibinfo {year} {2021})}\BibitemShut {NoStop}%
\bibitem [{\citenamefont {Tambasco}\ \emph {et~al.}(2018)\citenamefont
  {Tambasco}, \citenamefont {Corrielli}, \citenamefont {Chapman}, \citenamefont
  {Crespi}, \citenamefont {Zilberberg}, \citenamefont {Osellame},\ and\
  \citenamefont {Peruzzo}}]{tambasco2018quantum}%
  \BibitemOpen
  \bibfield  {author} {\bibinfo {author} {\bibfnamefont {J.-L.}\ \bibnamefont
  {Tambasco}}, \bibinfo {author} {\bibfnamefont {G.}~\bibnamefont {Corrielli}},
  \bibinfo {author} {\bibfnamefont {R.~J.}\ \bibnamefont {Chapman}}, \bibinfo
  {author} {\bibfnamefont {A.}~\bibnamefont {Crespi}}, \bibinfo {author}
  {\bibfnamefont {O.}~\bibnamefont {Zilberberg}}, \bibinfo {author}
  {\bibfnamefont {R.}~\bibnamefont {Osellame}},\ and\ \bibinfo {author}
  {\bibfnamefont {A.}~\bibnamefont {Peruzzo}},\ }\bibfield  {title} {\bibinfo
  {title} {Quantum interference of topological states of light},\ }\href
  {https://www.science.org/doi/10.1126/sciadv.aat3187} {\bibfield  {journal}
  {\bibinfo  {journal} {Sci. Adv.}\ }\textbf {\bibinfo {volume} {4}},\ \bibinfo
  {pages} {eaat3187} (\bibinfo {year} {2018})}\BibitemShut {NoStop}%
\bibitem [{\citenamefont {de~Vries}\ \emph {et~al.}(2018)\citenamefont
  {de~Vries}, \citenamefont {Timmerman}, \citenamefont {Ostroukh},
  \citenamefont {van Veen}, \citenamefont {Beukman}, \citenamefont {Qu},
  \citenamefont {Wimmer}, \citenamefont {Nguyen}, \citenamefont {Kiselev},
  \citenamefont {Yi}, \citenamefont {Sokolich}, \citenamefont {Manfra},
  \citenamefont {Marcus},\ and\ \citenamefont
  {Kouwenhoven}}]{PhysRevLett.120.047702}%
  \BibitemOpen
  \bibfield  {author} {\bibinfo {author} {\bibfnamefont {F.~K.}\ \bibnamefont
  {de~Vries}}, \bibinfo {author} {\bibfnamefont {T.}~\bibnamefont {Timmerman}},
  \bibinfo {author} {\bibfnamefont {V.~P.}\ \bibnamefont {Ostroukh}}, \bibinfo
  {author} {\bibfnamefont {J.}~\bibnamefont {van Veen}}, \bibinfo {author}
  {\bibfnamefont {A.~J.~A.}\ \bibnamefont {Beukman}}, \bibinfo {author}
  {\bibfnamefont {F.}~\bibnamefont {Qu}}, \bibinfo {author} {\bibfnamefont
  {M.}~\bibnamefont {Wimmer}}, \bibinfo {author} {\bibfnamefont {B.-M.}\
  \bibnamefont {Nguyen}}, \bibinfo {author} {\bibfnamefont {A.~A.}\
  \bibnamefont {Kiselev}}, \bibinfo {author} {\bibfnamefont {W.}~\bibnamefont
  {Yi}}, \bibinfo {author} {\bibfnamefont {M.}~\bibnamefont {Sokolich}},
  \bibinfo {author} {\bibfnamefont {M.~J.}\ \bibnamefont {Manfra}}, \bibinfo
  {author} {\bibfnamefont {C.~M.}\ \bibnamefont {Marcus}},\ and\ \bibinfo
  {author} {\bibfnamefont {L.~P.}\ \bibnamefont {Kouwenhoven}},\ }\bibfield
  {title} {\bibinfo {title} {$h/e$ superconducting quantum interference through
  trivial edge states in inas},\ }\href
  {https://doi.org/10.1103/PhysRevLett.120.047702} {\bibfield  {journal}
  {\bibinfo  {journal} {Phys. Rev. Lett.}\ }\textbf {\bibinfo {volume} {120}},\
  \bibinfo {pages} {047702} (\bibinfo {year} {2018})}\BibitemShut {NoStop}%
\bibitem [{\citenamefont {Bardarson}\ and\ \citenamefont
  {Moore}(2013)}]{bardarson2013quantum}%
  \BibitemOpen
  \bibfield  {author} {\bibinfo {author} {\bibfnamefont {J.~H.}\ \bibnamefont
  {Bardarson}}\ and\ \bibinfo {author} {\bibfnamefont {J.~E.}\ \bibnamefont
  {Moore}},\ }\bibfield  {title} {\bibinfo {title} {Quantum interference and
  {Aharonov--Bohm} oscillations in topological insulators},\ }\href
  {https://iopscience.iop.org/article/10.1088/0034-4885/76/5/056501} {\bibfield
   {journal} {\bibinfo  {journal} {Rep. Prog. Phys.}\ }\textbf {\bibinfo
  {volume} {76}},\ \bibinfo {pages} {056501} (\bibinfo {year}
  {2013})}\BibitemShut {NoStop}%
\bibitem [{\citenamefont {Hu}\ \emph {et~al.}(2020)\citenamefont {Hu},
  \citenamefont {Ke},\ and\ \citenamefont {Lee}}]{PhysRevA.101.052323}%
  \BibitemOpen
  \bibfield  {author} {\bibinfo {author} {\bibfnamefont {S.}~\bibnamefont
  {Hu}}, \bibinfo {author} {\bibfnamefont {Y.}~\bibnamefont {Ke}},\ and\
  \bibinfo {author} {\bibfnamefont {C.}~\bibnamefont {Lee}},\ }\bibfield
  {title} {\bibinfo {title} {Topological quantum transport and spatial
  entanglement distribution via a disordered bulk channel},\ }\href
  {https://doi.org/10.1103/PhysRevA.101.052323} {\bibfield  {journal} {\bibinfo
   {journal} {Phys. Rev. A}\ }\textbf {\bibinfo {volume} {101}},\ \bibinfo
  {pages} {052323} (\bibinfo {year} {2020})}\BibitemShut {NoStop}%
\bibitem [{\citenamefont {He}\ \emph {et~al.}(2019)\citenamefont {He},
  \citenamefont {Zhang},\ and\ \citenamefont {Zhang}}]{He:19}%
  \BibitemOpen
  \bibfield  {author} {\bibinfo {author} {\bibfnamefont {L.}~\bibnamefont
  {He}}, \bibinfo {author} {\bibfnamefont {W.~X.}\ \bibnamefont {Zhang}},\ and\
  \bibinfo {author} {\bibfnamefont {X.~D.}\ \bibnamefont {Zhang}},\ }\bibfield
  {title} {\bibinfo {title} {Topological all-optical logic gates based on
  two-dimensional photonic crystals},\ }\href
  {https://doi.org/10.1364/OE.27.025841} {\bibfield  {journal} {\bibinfo
  {journal} {Opt. Express}\ }\textbf {\bibinfo {volume} {27}},\ \bibinfo
  {pages} {25841} (\bibinfo {year} {2019})}\BibitemShut {NoStop}%
\bibitem [{\citenamefont {Boross}\ \emph {et~al.}(2019)\citenamefont {Boross},
  \citenamefont {Asb\'oth}, \citenamefont {Sz\'echenyi}, \citenamefont
  {Oroszl\'any},\ and\ \citenamefont {P\'alyi}}]{PhysRevB.100.045414}%
  \BibitemOpen
  \bibfield  {author} {\bibinfo {author} {\bibfnamefont {P.}~\bibnamefont
  {Boross}}, \bibinfo {author} {\bibfnamefont {J.~K.}\ \bibnamefont
  {Asb\'oth}}, \bibinfo {author} {\bibfnamefont {G.}~\bibnamefont
  {Sz\'echenyi}}, \bibinfo {author} {\bibfnamefont {L.}~\bibnamefont
  {Oroszl\'any}},\ and\ \bibinfo {author} {\bibfnamefont {A.}~\bibnamefont
  {P\'alyi}},\ }\bibfield  {title} {\bibinfo {title} {Poor man's topological
  quantum gate based on the su-schrieffer-heeger model},\ }\href
  {https://doi.org/10.1103/PhysRevB.100.045414} {\bibfield  {journal} {\bibinfo
   {journal} {Phys. Rev. B}\ }\textbf {\bibinfo {volume} {100}},\ \bibinfo
  {pages} {045414} (\bibinfo {year} {2019})}\BibitemShut {NoStop}%
\bibitem [{\citenamefont {He}\ \emph {et~al.}(2020)\citenamefont {He},
  \citenamefont {Ji}, \citenamefont {Wang},\ and\ \citenamefont
  {Zhang}}]{He:20}%
  \BibitemOpen
  \bibfield  {author} {\bibinfo {author} {\bibfnamefont {L.}~\bibnamefont
  {He}}, \bibinfo {author} {\bibfnamefont {H.~Y.}\ \bibnamefont {Ji}}, \bibinfo
  {author} {\bibfnamefont {Y.~J.}\ \bibnamefont {Wang}},\ and\ \bibinfo
  {author} {\bibfnamefont {X.~D.}\ \bibnamefont {Zhang}},\ }\bibfield  {title}
  {\bibinfo {title} {Topologically protected beam splitters and logic gates
  based on two-dimensional silicon photonic crystal slabs},\ }\href
  {https://doi.org/10.1364/OE.409265} {\bibfield  {journal} {\bibinfo
  {journal} {Opt. Express}\ }\textbf {\bibinfo {volume} {28}},\ \bibinfo
  {pages} {34015} (\bibinfo {year} {2020})}\BibitemShut {NoStop}%
\bibitem [{\citenamefont {Chao}\ \emph {et~al.}(2021)\citenamefont {Chao},
  \citenamefont {Cheng}, \citenamefont {Liu}, \citenamefont {Zhang},
  \citenamefont {Xu},\ and\ \citenamefont {Song}}]{chao2021novel}%
  \BibitemOpen
  \bibfield  {author} {\bibinfo {author} {\bibfnamefont {M.-H.}\ \bibnamefont
  {Chao}}, \bibinfo {author} {\bibfnamefont {B.}~\bibnamefont {Cheng}},
  \bibinfo {author} {\bibfnamefont {Q.-S.}\ \bibnamefont {Liu}}, \bibinfo
  {author} {\bibfnamefont {W.-J.}\ \bibnamefont {Zhang}}, \bibinfo {author}
  {\bibfnamefont {Y.}~\bibnamefont {Xu}},\ and\ \bibinfo {author}
  {\bibfnamefont {G.-F.}\ \bibnamefont {Song}},\ }\bibfield  {title} {\bibinfo
  {title} {Novel optical xor/or logic gates based on topologically protected
  valley photonic crystals edges},\ }\href
  {https://iopscience.iop.org/article/10.1088/2040-8986/ac11ac} {\bibfield
  {journal} {\bibinfo  {journal} {J. Opt.}\ }\textbf {\bibinfo {volume} {23}},\
  \bibinfo {pages} {115002} (\bibinfo {year} {2021})}\BibitemShut {NoStop}%
\bibitem [{\citenamefont {Naro{\.z}niak}\ \emph {et~al.}(2021)\citenamefont
  {Naro{\.z}niak}, \citenamefont {Dartiailh}, \citenamefont {Dowling},
  \citenamefont {Shabani},\ and\ \citenamefont
  {Byrnes}}]{narozniak2021quantum}%
  \BibitemOpen
  \bibfield  {author} {\bibinfo {author} {\bibfnamefont {M.}~\bibnamefont
  {Naro{\.z}niak}}, \bibinfo {author} {\bibfnamefont {M.~C.}\ \bibnamefont
  {Dartiailh}}, \bibinfo {author} {\bibfnamefont {J.~P.}\ \bibnamefont
  {Dowling}}, \bibinfo {author} {\bibfnamefont {J.}~\bibnamefont {Shabani}},\
  and\ \bibinfo {author} {\bibfnamefont {T.}~\bibnamefont {Byrnes}},\
  }\bibfield  {title} {\bibinfo {title} {Quantum gates for majoranas zero modes
  in topological superconductors in one-dimensional geometry},\ }\href
  {https://journals.aps.org/prb/abstract/10.1103/PhysRevB.103.205429}
  {\bibfield  {journal} {\bibinfo  {journal} {Phys. Rev. B}\ }\textbf {\bibinfo
  {volume} {103}},\ \bibinfo {pages} {205429} (\bibinfo {year}
  {2021})}\BibitemShut {NoStop}%
\bibitem [{\citenamefont {Ke}\ \emph {et~al.}(2017)\citenamefont {Ke},
  \citenamefont {Qin}, \citenamefont {Kivshar},\ and\ \citenamefont
  {Lee}}]{PhysRevA.95.063630}%
  \BibitemOpen
  \bibfield  {author} {\bibinfo {author} {\bibfnamefont {Y.}~\bibnamefont
  {Ke}}, \bibinfo {author} {\bibfnamefont {X.}~\bibnamefont {Qin}}, \bibinfo
  {author} {\bibfnamefont {Y.~S.}\ \bibnamefont {Kivshar}},\ and\ \bibinfo
  {author} {\bibfnamefont {C.}~\bibnamefont {Lee}},\ }\bibfield  {title}
  {\bibinfo {title} {Multiparticle wannier states and {Thouless} pumping of
  interacting bosons},\ }\href {https://doi.org/10.1103/PhysRevA.95.063630}
  {\bibfield  {journal} {\bibinfo  {journal} {Phys. Rev. A}\ }\textbf {\bibinfo
  {volume} {95}},\ \bibinfo {pages} {063630} (\bibinfo {year}
  {2017})}\BibitemShut {NoStop}%
\bibitem [{\citenamefont {Van~Voorden}\ and\ \citenamefont
  {Schoutens}(2019)}]{van2019topological}%
  \BibitemOpen
  \bibfield  {author} {\bibinfo {author} {\bibfnamefont {B.}~\bibnamefont
  {Van~Voorden}}\ and\ \bibinfo {author} {\bibfnamefont {K.}~\bibnamefont
  {Schoutens}},\ }\bibfield  {title} {\bibinfo {title} {Topological quantum
  pump of strongly interacting fermions in coupled chains},\ }\href
  {https://iopscience.iop.org/article/10.1088/1367-2630/aaf748} {\bibfield
  {journal} {\bibinfo  {journal} {New J. Phys.}\ }\textbf {\bibinfo {volume}
  {21}},\ \bibinfo {pages} {013026} (\bibinfo {year} {2019})}\BibitemShut
  {NoStop}%
\bibitem [{\citenamefont {Liu}\ \emph {et~al.}(2023{\natexlab{a}})\citenamefont
  {Liu}, \citenamefont {Hu}, \citenamefont {Zhang}, \citenamefont {Ke},\ and\
  \citenamefont {Lee}}]{PhysRevResearch.5.013020}%
  \BibitemOpen
  \bibfield  {author} {\bibinfo {author} {\bibfnamefont {W.}~\bibnamefont
  {Liu}}, \bibinfo {author} {\bibfnamefont {S.}~\bibnamefont {Hu}}, \bibinfo
  {author} {\bibfnamefont {L.}~\bibnamefont {Zhang}}, \bibinfo {author}
  {\bibfnamefont {Y.}~\bibnamefont {Ke}},\ and\ \bibinfo {author}
  {\bibfnamefont {C.}~\bibnamefont {Lee}},\ }\bibfield  {title} {\bibinfo
  {title} {Correlated topological pumping of interacting bosons assisted by
  bloch oscillations},\ }\href
  {https://doi.org/10.1103/PhysRevResearch.5.013020} {\bibfield  {journal}
  {\bibinfo  {journal} {Phys. Rev. Res.}\ }\textbf {\bibinfo {volume} {5}},\
  \bibinfo {pages} {013020} (\bibinfo {year} {2023}{\natexlab{a}})}\BibitemShut
  {NoStop}%
\bibitem [{\citenamefont {Lin}\ \emph {et~al.}(2020)\citenamefont {Lin},
  \citenamefont {Ke},\ and\ \citenamefont {Lee}}]{PhysRevA.101.023620}%
  \BibitemOpen
  \bibfield  {author} {\bibinfo {author} {\bibfnamefont {L.}~\bibnamefont
  {Lin}}, \bibinfo {author} {\bibfnamefont {Y.}~\bibnamefont {Ke}},\ and\
  \bibinfo {author} {\bibfnamefont {C.}~\bibnamefont {Lee}},\ }\bibfield
  {title} {\bibinfo {title} {Interaction-induced topological bound states and
  {Thouless} pumping in a one-dimensional optical lattice},\ }\href
  {https://doi.org/10.1103/PhysRevA.101.023620} {\bibfield  {journal} {\bibinfo
   {journal} {Phys. Rev. A}\ }\textbf {\bibinfo {volume} {101}},\ \bibinfo
  {pages} {023620} (\bibinfo {year} {2020})}\BibitemShut {NoStop}%
\bibitem [{\citenamefont {Huang}\ \emph {et~al.}(2024)\citenamefont {Huang},
  \citenamefont {Ke}, \citenamefont {Liu},\ and\ \citenamefont
  {Lee}}]{huang2024topological}%
  \BibitemOpen
  \bibfield  {author} {\bibinfo {author} {\bibfnamefont {B.}~\bibnamefont
  {Huang}}, \bibinfo {author} {\bibfnamefont {Y.}~\bibnamefont {Ke}}, \bibinfo
  {author} {\bibfnamefont {W.}~\bibnamefont {Liu}},\ and\ \bibinfo {author}
  {\bibfnamefont {C.}~\bibnamefont {Lee}},\ }\bibfield  {title} {\bibinfo
  {title} {Topological pumping induced by spatiotemporal modulation of
  interaction},\ }\href
  {https://iopscience.iop.org/article/10.1088/1402-4896/ad491e/meta} {\bibfield
   {journal} {\bibinfo  {journal} {Phys. Scr.}\ }\textbf {\bibinfo {volume}
  {99}},\ \bibinfo {pages} {065997} (\bibinfo {year} {2024})}\BibitemShut
  {NoStop}%
\bibitem [{\citenamefont {Walter}\ \emph {et~al.}(2023)\citenamefont {Walter},
  \citenamefont {Zhu}, \citenamefont {G{\"a}chter}, \citenamefont {Minguzzi},
  \citenamefont {Roschinski}, \citenamefont {Sandholzer}, \citenamefont
  {Viebahn},\ and\ \citenamefont {Esslinger}}]{walter2023quantization}%
  \BibitemOpen
  \bibfield  {author} {\bibinfo {author} {\bibfnamefont {A.-S.}\ \bibnamefont
  {Walter}}, \bibinfo {author} {\bibfnamefont {Z.}~\bibnamefont {Zhu}},
  \bibinfo {author} {\bibfnamefont {M.}~\bibnamefont {G{\"a}chter}}, \bibinfo
  {author} {\bibfnamefont {J.}~\bibnamefont {Minguzzi}}, \bibinfo {author}
  {\bibfnamefont {S.}~\bibnamefont {Roschinski}}, \bibinfo {author}
  {\bibfnamefont {K.}~\bibnamefont {Sandholzer}}, \bibinfo {author}
  {\bibfnamefont {K.}~\bibnamefont {Viebahn}},\ and\ \bibinfo {author}
  {\bibfnamefont {T.}~\bibnamefont {Esslinger}},\ }\bibfield  {title} {\bibinfo
  {title} {Quantization and its breakdown in a {Hubbard}--{Thouless} pump},\
  }\href {https://www.nature.com/articles/s41567-023-02145-w} {\bibfield
  {journal} {\bibinfo  {journal} {Nat. Phys.}\ }\textbf {\bibinfo {volume}
  {19}},\ \bibinfo {pages} {1471} (\bibinfo {year} {2023})}\BibitemShut
  {NoStop}%
\bibitem [{\citenamefont {Ke}\ and\ \citenamefont
  {Lee}(2023)}]{ke2023topological}%
  \BibitemOpen
  \bibfield  {author} {\bibinfo {author} {\bibfnamefont {Y.}~\bibnamefont
  {Ke}}\ and\ \bibinfo {author} {\bibfnamefont {C.}~\bibnamefont {Lee}},\
  }\bibfield  {title} {\bibinfo {title} {Topological quantum tango},\ }\href
  {https://www.nature.com/articles/s41567-023-02169-2} {\bibfield  {journal}
  {\bibinfo  {journal} {Nat. Phys.}\ }\textbf {\bibinfo {volume} {19}},\
  \bibinfo {pages} {1387} (\bibinfo {year} {2023})}\BibitemShut {NoStop}%
\bibitem [{\citenamefont {Bardyn}\ \emph {et~al.}(2016)\citenamefont {Bardyn},
  \citenamefont {Karzig}, \citenamefont {Refael},\ and\ \citenamefont
  {Liew}}]{PhysRevB.93.020502}%
  \BibitemOpen
  \bibfield  {author} {\bibinfo {author} {\bibfnamefont {C.-E.}\ \bibnamefont
  {Bardyn}}, \bibinfo {author} {\bibfnamefont {T.}~\bibnamefont {Karzig}},
  \bibinfo {author} {\bibfnamefont {G.}~\bibnamefont {Refael}},\ and\ \bibinfo
  {author} {\bibfnamefont {T.~C.~H.}\ \bibnamefont {Liew}},\ }\bibfield
  {title} {\bibinfo {title} {Chiral {Bogoliubov} excitations in nonlinear
  bosonic systems},\ }\href {https://doi.org/10.1103/PhysRevB.93.020502}
  {\bibfield  {journal} {\bibinfo  {journal} {Phys. Rev. B}\ }\textbf {\bibinfo
  {volume} {93}},\ \bibinfo {pages} {020502} (\bibinfo {year}
  {2016})}\BibitemShut {NoStop}%
\bibitem [{\citenamefont {Gulevich}\ \emph {et~al.}(2017)\citenamefont
  {Gulevich}, \citenamefont {Yudin}, \citenamefont {Skryabin}, \citenamefont
  {Iorsh},\ and\ \citenamefont {Shelykh}}]{gulevich2017exploring}%
  \BibitemOpen
  \bibfield  {author} {\bibinfo {author} {\bibfnamefont {D.~R.}\ \bibnamefont
  {Gulevich}}, \bibinfo {author} {\bibfnamefont {D.}~\bibnamefont {Yudin}},
  \bibinfo {author} {\bibfnamefont {D.~V.}\ \bibnamefont {Skryabin}}, \bibinfo
  {author} {\bibfnamefont {I.~V.}\ \bibnamefont {Iorsh}},\ and\ \bibinfo
  {author} {\bibfnamefont {I.~A.}\ \bibnamefont {Shelykh}},\ }\bibfield
  {title} {\bibinfo {title} {Exploring nonlinear topological states of matter
  with exciton-polaritons: Edge solitons in kagome lattice},\ }\href
  {https://www.nature.com/articles/s41598-017-01646-y#citeas} {\bibfield
  {journal} {\bibinfo  {journal} {Sci. Rep.}\ }\textbf {\bibinfo {volume}
  {7}},\ \bibinfo {pages} {1780} (\bibinfo {year} {2017})}\BibitemShut
  {NoStop}%
\bibitem [{\citenamefont {Leykam}\ and\ \citenamefont
  {Chong}(2016)}]{leykam2016edge}%
  \BibitemOpen
  \bibfield  {author} {\bibinfo {author} {\bibfnamefont {D.}~\bibnamefont
  {Leykam}}\ and\ \bibinfo {author} {\bibfnamefont {Y.~D.}\ \bibnamefont
  {Chong}},\ }\bibfield  {title} {\bibinfo {title} {Edge solitons in
  nonlinear-photonic topological insulators},\ }\href
  {https://journals.aps.org/prl/abstract/10.1103/PhysRevLett.117.143901}
  {\bibfield  {journal} {\bibinfo  {journal} {Phys. Rev. Lett.}\ }\textbf
  {\bibinfo {volume} {117}},\ \bibinfo {pages} {143901} (\bibinfo {year}
  {2016})}\BibitemShut {NoStop}%
\bibitem [{\citenamefont {Bleu}\ \emph {et~al.}(2016)\citenamefont {Bleu},
  \citenamefont {Solnyshkov},\ and\ \citenamefont
  {Malpuech}}]{bleu2016interacting}%
  \BibitemOpen
  \bibfield  {author} {\bibinfo {author} {\bibfnamefont {O.}~\bibnamefont
  {Bleu}}, \bibinfo {author} {\bibfnamefont {D.}~\bibnamefont {Solnyshkov}},\
  and\ \bibinfo {author} {\bibfnamefont {G.}~\bibnamefont {Malpuech}},\
  }\bibfield  {title} {\bibinfo {title} {Interacting quantum fluid in a
  polariton {Chern} insulator},\ }\href
  {https://doi.org/10.1103/PhysRevB.93.085438} {\bibfield  {journal} {\bibinfo
  {journal} {Phys. Rev. B}\ }\textbf {\bibinfo {volume} {93}},\ \bibinfo
  {pages} {085438} (\bibinfo {year} {2016})}\BibitemShut {NoStop}%
\bibitem [{\citenamefont {Bleu}\ \emph {et~al.}(2017)\citenamefont {Bleu},
  \citenamefont {Solnyshkov},\ and\ \citenamefont
  {Malpuech}}]{bleu2017photonic}%
  \BibitemOpen
  \bibfield  {author} {\bibinfo {author} {\bibfnamefont {O.}~\bibnamefont
  {Bleu}}, \bibinfo {author} {\bibfnamefont {D.}~\bibnamefont {Solnyshkov}},\
  and\ \bibinfo {author} {\bibfnamefont {G.}~\bibnamefont {Malpuech}},\
  }\bibfield  {title} {\bibinfo {title} {Photonic versus electronic quantum
  anomalous {Hall} effect},\ }\href
  {https://journals.aps.org/prb/abstract/10.1103/PhysRevB.95.115415} {\bibfield
   {journal} {\bibinfo  {journal} {Phys. Rev. B}\ }\textbf {\bibinfo {volume}
  {95}},\ \bibinfo {pages} {115415} (\bibinfo {year} {2017})}\BibitemShut
  {NoStop}%
\bibitem [{\citenamefont {Liu}\ \emph {et~al.}(2017)\citenamefont {Liu},
  \citenamefont {Wang}, \citenamefont {Yin}, \citenamefont {Wu}, \citenamefont
  {Xu}, \citenamefont {Wen},\ and\ \citenamefont {Chen}}]{LIU2017183}%
  \BibitemOpen
  \bibfield  {author} {\bibinfo {author} {\bibfnamefont {C.}~\bibnamefont
  {Liu}}, \bibinfo {author} {\bibfnamefont {Z.}~\bibnamefont {Wang}}, \bibinfo
  {author} {\bibfnamefont {C.}~\bibnamefont {Yin}}, \bibinfo {author}
  {\bibfnamefont {Y.}~\bibnamefont {Wu}}, \bibinfo {author} {\bibfnamefont
  {T.}~\bibnamefont {Xu}}, \bibinfo {author} {\bibfnamefont {L.}~\bibnamefont
  {Wen}},\ and\ \bibinfo {author} {\bibfnamefont {S.}~\bibnamefont {Chen}},\
  }\bibfield  {title} {\bibinfo {title} {The nontrivial states in
  one-dimensional nonlinear bichromatic superlattices},\ }\href
  {https://doi.org/https://doi.org/10.1016/j.physe.2017.01.033} {\bibfield
  {journal} {\bibinfo  {journal} {Physica E: Low Dimens. Syst. Nanostruct.}\
  }\textbf {\bibinfo {volume} {90}},\ \bibinfo {pages} {183} (\bibinfo {year}
  {2017})}\BibitemShut {NoStop}%
\bibitem [{\citenamefont {Gross}(1961)}]{gross1961structure}%
  \BibitemOpen
  \bibfield  {author} {\bibinfo {author} {\bibfnamefont {E.~P.}\ \bibnamefont
  {Gross}},\ }\bibfield  {title} {\bibinfo {title} {Structure of a quantized
  vortex in boson systems},\ }\href
  {https://link.springer.com/article/10.1007/BF02731494} {\bibfield  {journal}
  {\bibinfo  {journal} {Il Nuovo Cimento (1955-1965)}\ }\textbf {\bibinfo
  {volume} {20}},\ \bibinfo {pages} {454} (\bibinfo {year} {1961})}\BibitemShut
  {NoStop}%
\bibitem [{\citenamefont {Pitaevskii}(1961)}]{pitaevskii1961vortex}%
  \BibitemOpen
  \bibfield  {author} {\bibinfo {author} {\bibfnamefont {L.~P.}\ \bibnamefont
  {Pitaevskii}},\ }\bibfield  {title} {\bibinfo {title} {Vortex lines in an
  imperfect {Bose} gas},\ }\href
  {https://www.jetp.ras.ru/cgi-bin/dn/e_013_02_0451.pdf} {\bibfield  {journal}
  {\bibinfo  {journal} {Sov. Phys. JETP}\ }\textbf {\bibinfo {volume} {13}},\
  \bibinfo {pages} {451} (\bibinfo {year} {1961})}\BibitemShut {NoStop}%
\bibitem [{\citenamefont {J\"urgensen}\ and\ \citenamefont
  {Rechtsman}(2022)}]{PhysRevLett.128.113901}%
  \BibitemOpen
  \bibfield  {author} {\bibinfo {author} {\bibfnamefont {M.}~\bibnamefont
  {J\"urgensen}}\ and\ \bibinfo {author} {\bibfnamefont {M.~C.}\ \bibnamefont
  {Rechtsman}},\ }\bibfield  {title} {\bibinfo {title} {{Chern} number governs
  soliton motion in nonlinear {Thouless} pumps},\ }\href
  {https://doi.org/10.1103/PhysRevLett.128.113901} {\bibfield  {journal}
  {\bibinfo  {journal} {Phys. Rev. Lett.}\ }\textbf {\bibinfo {volume} {128}},\
  \bibinfo {pages} {113901} (\bibinfo {year} {2022})}\BibitemShut {NoStop}%
\bibitem [{\citenamefont {Mostaan}\ \emph {et~al.}(2022)\citenamefont
  {Mostaan}, \citenamefont {Grusdt},\ and\ \citenamefont
  {Goldman}}]{mostaan2022quantized}%
  \BibitemOpen
  \bibfield  {author} {\bibinfo {author} {\bibfnamefont {N.}~\bibnamefont
  {Mostaan}}, \bibinfo {author} {\bibfnamefont {F.}~\bibnamefont {Grusdt}},\
  and\ \bibinfo {author} {\bibfnamefont {N.}~\bibnamefont {Goldman}},\
  }\bibfield  {title} {\bibinfo {title} {Quantized topological pumping of
  solitons in nonlinear photonics and ultracold atomic mixtures},\ }\href
  {https://www.nature.com/articles/s41467-022-33478-4} {\bibfield  {journal}
  {\bibinfo  {journal} {Nat. Commun.}\ }\textbf {\bibinfo {volume} {13}},\
  \bibinfo {pages} {5997} (\bibinfo {year} {2022})}\BibitemShut {NoStop}%
\bibitem [{\citenamefont {Fu}\ \emph {et~al.}(2022{\natexlab{a}})\citenamefont
  {Fu}, \citenamefont {Wang}, \citenamefont {Kartashov}, \citenamefont
  {Konotop},\ and\ \citenamefont {Ye}}]{PhysRevLett.128.154101}%
  \BibitemOpen
  \bibfield  {author} {\bibinfo {author} {\bibfnamefont {Q.}~\bibnamefont
  {Fu}}, \bibinfo {author} {\bibfnamefont {P.}~\bibnamefont {Wang}}, \bibinfo
  {author} {\bibfnamefont {Y.~V.}\ \bibnamefont {Kartashov}}, \bibinfo {author}
  {\bibfnamefont {V.~V.}\ \bibnamefont {Konotop}},\ and\ \bibinfo {author}
  {\bibfnamefont {F.}~\bibnamefont {Ye}},\ }\bibfield  {title} {\bibinfo
  {title} {Nonlinear {Thouless} pumping: Solitons and transport breakdown},\
  }\href {https://doi.org/10.1103/PhysRevLett.128.154101} {\bibfield  {journal}
  {\bibinfo  {journal} {Phys. Rev. Lett.}\ }\textbf {\bibinfo {volume} {128}},\
  \bibinfo {pages} {154101} (\bibinfo {year} {2022}{\natexlab{a}})}\BibitemShut
  {NoStop}%
\bibitem [{\citenamefont {Fu}\ \emph {et~al.}(2022{\natexlab{b}})\citenamefont
  {Fu}, \citenamefont {Wang}, \citenamefont {Kartashov}, \citenamefont
  {Konotop},\ and\ \citenamefont {Ye}}]{PhysRevLett.129.183901}%
  \BibitemOpen
  \bibfield  {author} {\bibinfo {author} {\bibfnamefont {Q.}~\bibnamefont
  {Fu}}, \bibinfo {author} {\bibfnamefont {P.}~\bibnamefont {Wang}}, \bibinfo
  {author} {\bibfnamefont {Y.~V.}\ \bibnamefont {Kartashov}}, \bibinfo {author}
  {\bibfnamefont {V.~V.}\ \bibnamefont {Konotop}},\ and\ \bibinfo {author}
  {\bibfnamefont {F.}~\bibnamefont {Ye}},\ }\bibfield  {title} {\bibinfo
  {title} {Two-dimensional nonlinear {Thouless} pumping of matter waves},\
  }\href {https://doi.org/10.1103/PhysRevLett.129.183901} {\bibfield  {journal}
  {\bibinfo  {journal} {Phys. Rev. Lett.}\ }\textbf {\bibinfo {volume} {129}},\
  \bibinfo {pages} {183901} (\bibinfo {year} {2022}{\natexlab{b}})}\BibitemShut
  {NoStop}%
\bibitem [{\citenamefont {Tuloup}\ \emph {et~al.}(2023)\citenamefont {Tuloup},
  \citenamefont {Bomantara},\ and\ \citenamefont {Gong}}]{Tuloup_2023}%
  \BibitemOpen
  \bibfield  {author} {\bibinfo {author} {\bibfnamefont {T.}~\bibnamefont
  {Tuloup}}, \bibinfo {author} {\bibfnamefont {R.~W.}\ \bibnamefont
  {Bomantara}},\ and\ \bibinfo {author} {\bibfnamefont {J.}~\bibnamefont
  {Gong}},\ }\bibfield  {title} {\bibinfo {title} {Breakdown of quantization in
  nonlinear {Thouless} pumping},\ }\href
  {https://doi.org/10.1088/1367-2630/acef4d} {\bibfield  {journal} {\bibinfo
  {journal} {New J. Phys.}\ }\textbf {\bibinfo {volume} {25}},\ \bibinfo
  {pages} {083048} (\bibinfo {year} {2023})}\BibitemShut {NoStop}%
\bibitem [{\citenamefont {Yuan}\ \emph {et~al.}(2023)\citenamefont {Yuan},
  \citenamefont {Dai},\ and\ \citenamefont {Chen}}]{16-20230740}%
  \BibitemOpen
  \bibfield  {author} {\bibinfo {author} {\bibfnamefont {T.}~\bibnamefont
  {Yuan}}, \bibinfo {author} {\bibfnamefont {H.-N.}\ \bibnamefont {Dai}},\ and\
  \bibinfo {author} {\bibfnamefont {Y.-A.}\ \bibnamefont {Chen}},\ }\bibfield
  {title} {\bibinfo {title} {Nonlinear topological pumping in momentum space
  lattice of ultracold atoms},\ }\href
  {https://wulixb.iphy.ac.cn/en/article/doi/10.7498/aps.72.20230740} {\bibfield
   {journal} {\bibinfo  {journal} {ACTA PHYSICA SINICA}\ }\textbf {\bibinfo
  {volume} {72}} (\bibinfo {year} {2023})}\BibitemShut {NoStop}%
\bibitem [{\citenamefont {Lyu}\ \emph {et~al.}(2024)\citenamefont {Lyu},
  \citenamefont {Zhang},\ and\ \citenamefont
  {Busch}}]{PhysRevResearch.6.L042010}%
  \BibitemOpen
  \bibfield  {author} {\bibinfo {author} {\bibfnamefont {H.}~\bibnamefont
  {Lyu}}, \bibinfo {author} {\bibfnamefont {Y.}~\bibnamefont {Zhang}},\ and\
  \bibinfo {author} {\bibfnamefont {T.}~\bibnamefont {Busch}},\ }\bibfield
  {title} {\bibinfo {title} {{Thouless} pumping and trapping of two-component
  gap solitons},\ }\href {https://doi.org/10.1103/PhysRevResearch.6.L042010}
  {\bibfield  {journal} {\bibinfo  {journal} {Phys. Rev. Res.}\ }\textbf
  {\bibinfo {volume} {6}},\ \bibinfo {pages} {L042010} (\bibinfo {year}
  {2024})}\BibitemShut {NoStop}%
\bibitem [{\citenamefont {J{\"u}rgensen}\ \emph {et~al.}(2021)\citenamefont
  {J{\"u}rgensen}, \citenamefont {Mukherjee},\ and\ \citenamefont
  {Rechtsman}}]{jurgensen2021quantized}%
  \BibitemOpen
  \bibfield  {author} {\bibinfo {author} {\bibfnamefont {M.}~\bibnamefont
  {J{\"u}rgensen}}, \bibinfo {author} {\bibfnamefont {S.}~\bibnamefont
  {Mukherjee}},\ and\ \bibinfo {author} {\bibfnamefont {M.~C.}\ \bibnamefont
  {Rechtsman}},\ }\bibfield  {title} {\bibinfo {title} {Quantized nonlinear
  {Thouless} pumping},\ }\href
  {https://www.nature.com/articles/s41586-021-03688-9} {\bibfield  {journal}
  {\bibinfo  {journal} {Nature}\ }\textbf {\bibinfo {volume} {596}},\ \bibinfo
  {pages} {63} (\bibinfo {year} {2021})}\BibitemShut {NoStop}%
\bibitem [{\citenamefont {J{\"u}rgensen}\ \emph {et~al.}(2023)\citenamefont
  {J{\"u}rgensen}, \citenamefont {Mukherjee}, \citenamefont {J{\"o}rg},\ and\
  \citenamefont {Rechtsman}}]{jurgensen2023quantized}%
  \BibitemOpen
  \bibfield  {author} {\bibinfo {author} {\bibfnamefont {M.}~\bibnamefont
  {J{\"u}rgensen}}, \bibinfo {author} {\bibfnamefont {S.}~\bibnamefont
  {Mukherjee}}, \bibinfo {author} {\bibfnamefont {C.}~\bibnamefont
  {J{\"o}rg}},\ and\ \bibinfo {author} {\bibfnamefont {M.~C.}\ \bibnamefont
  {Rechtsman}},\ }\bibfield  {title} {\bibinfo {title} {Quantized fractional
  {Thouless} pumping of solitons},\ }\href
  {https://www.nature.com/articles/s41586-021-03688-9} {\bibfield  {journal}
  {\bibinfo  {journal} {Nat. Phys.}\ }\textbf {\bibinfo {volume} {19}},\
  \bibinfo {pages} {420} (\bibinfo {year} {2023})}\BibitemShut {NoStop}%
\bibitem [{\citenamefont {Hatsugai}\ and\ \citenamefont
  {Fukui}(2016)}]{PhysRevB.94.041102}%
  \BibitemOpen
  \bibfield  {author} {\bibinfo {author} {\bibfnamefont {Y.}~\bibnamefont
  {Hatsugai}}\ and\ \bibinfo {author} {\bibfnamefont {T.}~\bibnamefont
  {Fukui}},\ }\bibfield  {title} {\bibinfo {title} {Bulk-edge correspondence in
  topological pumping},\ }\href {https://doi.org/10.1103/PhysRevB.94.041102}
  {\bibfield  {journal} {\bibinfo  {journal} {Phys. Rev. B}\ }\textbf {\bibinfo
  {volume} {94}},\ \bibinfo {pages} {041102} (\bibinfo {year}
  {2016})}\BibitemShut {NoStop}%
\bibitem [{\citenamefont {Bloch}(2005)}]{bloch2005ultracold}%
  \BibitemOpen
  \bibfield  {author} {\bibinfo {author} {\bibfnamefont {I.}~\bibnamefont
  {Bloch}},\ }\bibfield  {title} {\bibinfo {title} {Ultracold quantum gases in
  optical lattices},\ }\href {https://www.nature.com/articles/nphys138}
  {\bibfield  {journal} {\bibinfo  {journal} {Nat. Phys.}\ }\textbf {\bibinfo
  {volume} {1}},\ \bibinfo {pages} {23} (\bibinfo {year} {2005})}\BibitemShut
  {NoStop}%
\bibitem [{\citenamefont {Xu}\ \emph {et~al.}(2013)\citenamefont {Xu},
  \citenamefont {Li},\ and\ \citenamefont {Chen}}]{PhysRevLett.110.215301}%
  \BibitemOpen
  \bibfield  {author} {\bibinfo {author} {\bibfnamefont {Z.}~\bibnamefont
  {Xu}}, \bibinfo {author} {\bibfnamefont {L.}~\bibnamefont {Li}},\ and\
  \bibinfo {author} {\bibfnamefont {S.}~\bibnamefont {Chen}},\ }\bibfield
  {title} {\bibinfo {title} {Fractional topological states of dipolar fermions
  in one-dimensional optical superlattices},\ }\href
  {https://doi.org/10.1103/PhysRevLett.110.215301} {\bibfield  {journal}
  {\bibinfo  {journal} {Phys. Rev. Lett.}\ }\textbf {\bibinfo {volume} {110}},\
  \bibinfo {pages} {215301} (\bibinfo {year} {2013})}\BibitemShut {NoStop}%
\bibitem [{\citenamefont {Zhou}\ \emph {et~al.}(2015)\citenamefont {Zhou},
  \citenamefont {Tan},\ and\ \citenamefont {Gong}}]{PhysRevB.92.245409}%
  \BibitemOpen
  \bibfield  {author} {\bibinfo {author} {\bibfnamefont {L.}~\bibnamefont
  {Zhou}}, \bibinfo {author} {\bibfnamefont {D.~Y.}\ \bibnamefont {Tan}},\ and\
  \bibinfo {author} {\bibfnamefont {J.}~\bibnamefont {Gong}},\ }\bibfield
  {title} {\bibinfo {title} {Effects of dephasing on quantum adiabatic pumping
  with nonequilibrium initial states},\ }\href
  {https://doi.org/10.1103/PhysRevB.92.245409} {\bibfield  {journal} {\bibinfo
  {journal} {Phys. Rev. B}\ }\textbf {\bibinfo {volume} {92}},\ \bibinfo
  {pages} {245409} (\bibinfo {year} {2015})}\BibitemShut {NoStop}%
\bibitem [{\citenamefont {Zeng}\ \emph {et~al.}(2015)\citenamefont {Zeng},
  \citenamefont {Wang},\ and\ \citenamefont {Zhai}}]{PhysRevLett.115.095302}%
  \BibitemOpen
  \bibfield  {author} {\bibinfo {author} {\bibfnamefont {T.-S.}\ \bibnamefont
  {Zeng}}, \bibinfo {author} {\bibfnamefont {C.}~\bibnamefont {Wang}},\ and\
  \bibinfo {author} {\bibfnamefont {H.}~\bibnamefont {Zhai}},\ }\bibfield
  {title} {\bibinfo {title} {Charge pumping of interacting fermion atoms in the
  synthetic dimension},\ }\href
  {https://doi.org/10.1103/PhysRevLett.115.095302} {\bibfield  {journal}
  {\bibinfo  {journal} {Phys. Rev. Lett.}\ }\textbf {\bibinfo {volume} {115}},\
  \bibinfo {pages} {095302} (\bibinfo {year} {2015})}\BibitemShut {NoStop}%
\bibitem [{\citenamefont {Zeng}\ \emph {et~al.}(2016)\citenamefont {Zeng},
  \citenamefont {Zhu},\ and\ \citenamefont {Sheng}}]{PhysRevB.94.235139}%
  \BibitemOpen
  \bibfield  {author} {\bibinfo {author} {\bibfnamefont {T.-S.}\ \bibnamefont
  {Zeng}}, \bibinfo {author} {\bibfnamefont {W.}~\bibnamefont {Zhu}},\ and\
  \bibinfo {author} {\bibfnamefont {D.~N.}\ \bibnamefont {Sheng}},\ }\bibfield
  {title} {\bibinfo {title} {Fractional charge pumping of interacting bosons in
  one-dimensional superlattice},\ }\href
  {https://doi.org/10.1103/PhysRevB.94.235139} {\bibfield  {journal} {\bibinfo
  {journal} {Phys. Rev. B}\ }\textbf {\bibinfo {volume} {94}},\ \bibinfo
  {pages} {235139} (\bibinfo {year} {2016})}\BibitemShut {NoStop}%
\bibitem [{\citenamefont {Goldman}\ \emph {et~al.}(2016)\citenamefont
  {Goldman}, \citenamefont {Budich},\ and\ \citenamefont
  {Zoller}}]{goldman2016topological}%
  \BibitemOpen
  \bibfield  {author} {\bibinfo {author} {\bibfnamefont {N.}~\bibnamefont
  {Goldman}}, \bibinfo {author} {\bibfnamefont {J.~C.}\ \bibnamefont
  {Budich}},\ and\ \bibinfo {author} {\bibfnamefont {P.}~\bibnamefont
  {Zoller}},\ }\bibfield  {title} {\bibinfo {title} {Topological quantum matter
  with ultracold gases in optical lattices},\ }\href
  {https://www.nature.com/articles/nphys3803} {\bibfield  {journal} {\bibinfo
  {journal} {Nat. Phys.}\ }\textbf {\bibinfo {volume} {12}},\ \bibinfo {pages}
  {639} (\bibinfo {year} {2016})}\BibitemShut {NoStop}%
\bibitem [{\citenamefont {Tai}\ \emph {et~al.}(2017)\citenamefont {Tai},
  \citenamefont {Lukin}, \citenamefont {Rispoli}, \citenamefont {Schittko},
  \citenamefont {Menke}, \citenamefont {Borgnia}, \citenamefont {Preiss},
  \citenamefont {Grusdt}, \citenamefont {Kaufman},\ and\ \citenamefont
  {Greiner}}]{tai2017microscopy}%
  \BibitemOpen
  \bibfield  {author} {\bibinfo {author} {\bibfnamefont {M.~E.}\ \bibnamefont
  {Tai}}, \bibinfo {author} {\bibfnamefont {A.}~\bibnamefont {Lukin}}, \bibinfo
  {author} {\bibfnamefont {M.}~\bibnamefont {Rispoli}}, \bibinfo {author}
  {\bibfnamefont {R.}~\bibnamefont {Schittko}}, \bibinfo {author}
  {\bibfnamefont {T.}~\bibnamefont {Menke}}, \bibinfo {author} {\bibfnamefont
  {D.}~\bibnamefont {Borgnia}}, \bibinfo {author} {\bibfnamefont {P.~M.}\
  \bibnamefont {Preiss}}, \bibinfo {author} {\bibfnamefont {F.}~\bibnamefont
  {Grusdt}}, \bibinfo {author} {\bibfnamefont {A.~M.}\ \bibnamefont
  {Kaufman}},\ and\ \bibinfo {author} {\bibfnamefont {M.}~\bibnamefont
  {Greiner}},\ }\bibfield  {title} {\bibinfo {title} {Microscopy of the
  interacting harper--hofstadter model in the two-body limit},\ }\href
  {https://www.nature.com/articles/nature22811} {\bibfield  {journal} {\bibinfo
   {journal} {Nature}\ }\textbf {\bibinfo {volume} {546}},\ \bibinfo {pages}
  {519} (\bibinfo {year} {2017})}\BibitemShut {NoStop}%
\bibitem [{\citenamefont {Lindner}\ \emph {et~al.}(2017)\citenamefont
  {Lindner}, \citenamefont {Berg},\ and\ \citenamefont
  {Rudner}}]{lindner2017universal}%
  \BibitemOpen
  \bibfield  {author} {\bibinfo {author} {\bibfnamefont {N.~H.}\ \bibnamefont
  {Lindner}}, \bibinfo {author} {\bibfnamefont {E.}~\bibnamefont {Berg}},\ and\
  \bibinfo {author} {\bibfnamefont {M.~S.}\ \bibnamefont {Rudner}},\ }\bibfield
   {title} {\bibinfo {title} {Universal chiral quasisteady states in
  periodically driven many-body systems},\ }\href
  {https://journals.aps.org/prx/abstract/10.1103/PhysRevX.7.011018} {\bibfield
  {journal} {\bibinfo  {journal} {Phys. Rev. X}\ }\textbf {\bibinfo {volume}
  {7}},\ \bibinfo {pages} {011018} (\bibinfo {year} {2017})}\BibitemShut
  {NoStop}%
\bibitem [{\citenamefont {Rice}\ and\ \citenamefont
  {Mele}(1982)}]{PhysRevLett.49.1455}%
  \BibitemOpen
  \bibfield  {author} {\bibinfo {author} {\bibfnamefont {M.~J.}\ \bibnamefont
  {Rice}}\ and\ \bibinfo {author} {\bibfnamefont {E.~J.}\ \bibnamefont
  {Mele}},\ }\bibfield  {title} {\bibinfo {title} {Elementary excitations of a
  linearly conjugated diatomic polymer},\ }\href
  {https://doi.org/10.1103/PhysRevLett.49.1455} {\bibfield  {journal} {\bibinfo
   {journal} {Phys. Rev. Lett.}\ }\textbf {\bibinfo {volume} {49}},\ \bibinfo
  {pages} {1455} (\bibinfo {year} {1982})}\BibitemShut {NoStop}%
\bibitem [{\citenamefont {Liu}\ \emph {et~al.}(2023{\natexlab{b}})\citenamefont
  {Liu}, \citenamefont {Ke}, \citenamefont {Lei},\ and\ \citenamefont
  {Lee}}]{Liu_2023}%
  \BibitemOpen
  \bibfield  {author} {\bibinfo {author} {\bibfnamefont {W.}~\bibnamefont
  {Liu}}, \bibinfo {author} {\bibfnamefont {Y.}~\bibnamefont {Ke}}, \bibinfo
  {author} {\bibfnamefont {Z.}~\bibnamefont {Lei}},\ and\ \bibinfo {author}
  {\bibfnamefont {C.}~\bibnamefont {Lee}},\ }\bibfield  {title} {\bibinfo
  {title} {Magnon boundary states tailored by longitudinal spin–spin
  interactions and topology},\ }\href
  {https://doi.org/10.1088/1367-2630/acf8ea} {\bibfield  {journal} {\bibinfo
  {journal} {New Journal of Physics}\ }\textbf {\bibinfo {volume} {25}},\
  \bibinfo {pages} {093042} (\bibinfo {year} {2023}{\natexlab{b}})}\BibitemShut
  {NoStop}%
\bibitem [{\citenamefont {Oka}\ and\ \citenamefont
  {Aoki}(2009)}]{PhysRevB.79.081406}%
  \BibitemOpen
  \bibfield  {author} {\bibinfo {author} {\bibfnamefont {T.}~\bibnamefont
  {Oka}}\ and\ \bibinfo {author} {\bibfnamefont {H.}~\bibnamefont {Aoki}},\
  }\bibfield  {title} {\bibinfo {title} {Photovoltaic hall effect in
  graphene},\ }\href {https://doi.org/10.1103/PhysRevB.79.081406} {\bibfield
  {journal} {\bibinfo  {journal} {Phys. Rev. B}\ }\textbf {\bibinfo {volume}
  {79}},\ \bibinfo {pages} {081406} (\bibinfo {year} {2009})}\BibitemShut
  {NoStop}%
\bibitem [{\citenamefont {Kitagawa}\ \emph {et~al.}(2010)\citenamefont
  {Kitagawa}, \citenamefont {Berg}, \citenamefont {Rudner},\ and\ \citenamefont
  {Demler}}]{PhysRevB.82.235114}%
  \BibitemOpen
  \bibfield  {author} {\bibinfo {author} {\bibfnamefont {T.}~\bibnamefont
  {Kitagawa}}, \bibinfo {author} {\bibfnamefont {E.}~\bibnamefont {Berg}},
  \bibinfo {author} {\bibfnamefont {M.}~\bibnamefont {Rudner}},\ and\ \bibinfo
  {author} {\bibfnamefont {E.}~\bibnamefont {Demler}},\ }\bibfield  {title}
  {\bibinfo {title} {Topological characterization of periodically driven
  quantum systems},\ }\href {https://doi.org/10.1103/PhysRevB.82.235114}
  {\bibfield  {journal} {\bibinfo  {journal} {Phys. Rev. B}\ }\textbf {\bibinfo
  {volume} {82}},\ \bibinfo {pages} {235114} (\bibinfo {year}
  {2010})}\BibitemShut {NoStop}%
\bibitem [{\citenamefont {Lindner}\ \emph {et~al.}(2011)\citenamefont
  {Lindner}, \citenamefont {Refael},\ and\ \citenamefont
  {Galitski}}]{lindner2011floquet}%
  \BibitemOpen
  \bibfield  {author} {\bibinfo {author} {\bibfnamefont {N.~H.}\ \bibnamefont
  {Lindner}}, \bibinfo {author} {\bibfnamefont {G.}~\bibnamefont {Refael}},\
  and\ \bibinfo {author} {\bibfnamefont {V.}~\bibnamefont {Galitski}},\
  }\bibfield  {title} {\bibinfo {title} {Floquet topological insulator in
  semiconductor quantum wells},\ }\href
  {https://www.nature.com/articles/nphys1926} {\bibfield  {journal} {\bibinfo
  {journal} {Nat. Phys.}\ }\textbf {\bibinfo {volume} {7}},\ \bibinfo {pages}
  {490} (\bibinfo {year} {2011})}\BibitemShut {NoStop}%
\bibitem [{\citenamefont {Rechtsman}\ \emph {et~al.}(2013)\citenamefont
  {Rechtsman}, \citenamefont {Zeuner}, \citenamefont {Plotnik}, \citenamefont
  {Lumer}, \citenamefont {Podolsky}, \citenamefont {Dreisow}, \citenamefont
  {Nolte}, \citenamefont {Segev},\ and\ \citenamefont
  {Szameit}}]{rechtsman2013photonic}%
  \BibitemOpen
  \bibfield  {author} {\bibinfo {author} {\bibfnamefont {M.~C.}\ \bibnamefont
  {Rechtsman}}, \bibinfo {author} {\bibfnamefont {J.~M.}\ \bibnamefont
  {Zeuner}}, \bibinfo {author} {\bibfnamefont {Y.}~\bibnamefont {Plotnik}},
  \bibinfo {author} {\bibfnamefont {Y.}~\bibnamefont {Lumer}}, \bibinfo
  {author} {\bibfnamefont {D.}~\bibnamefont {Podolsky}}, \bibinfo {author}
  {\bibfnamefont {F.}~\bibnamefont {Dreisow}}, \bibinfo {author} {\bibfnamefont
  {S.}~\bibnamefont {Nolte}}, \bibinfo {author} {\bibfnamefont
  {M.}~\bibnamefont {Segev}},\ and\ \bibinfo {author} {\bibfnamefont
  {A.}~\bibnamefont {Szameit}},\ }\bibfield  {title} {\bibinfo {title}
  {Photonic {Floquet} topological insulators},\ }\href
  {https://www.nature.com/articles/nature12066} {\bibfield  {journal} {\bibinfo
   {journal} {Nature}\ }\textbf {\bibinfo {volume} {496}},\ \bibinfo {pages}
  {196} (\bibinfo {year} {2013})}\BibitemShut {NoStop}%
\bibitem [{\citenamefont {Zhu}\ \emph {et~al.}(2024)\citenamefont {Zhu},
  \citenamefont {Tan}, \citenamefont {Ke},\ and\ \citenamefont
  {Zhong}}]{PhysRevB.109.224315}%
  \BibitemOpen
  \bibfield  {author} {\bibinfo {author} {\bibfnamefont {B.}~\bibnamefont
  {Zhu}}, \bibinfo {author} {\bibfnamefont {Z.}~\bibnamefont {Tan}}, \bibinfo
  {author} {\bibfnamefont {Y.}~\bibnamefont {Ke}},\ and\ \bibinfo {author}
  {\bibfnamefont {H.}~\bibnamefont {Zhong}},\ }\bibfield  {title} {\bibinfo
  {title} {Dynamic winding number for {Floquet} topological insulators with
  arbitrarily driving frequencies},\ }\href
  {https://doi.org/10.1103/PhysRevB.109.224315} {\bibfield  {journal} {\bibinfo
   {journal} {Phys. Rev. B}\ }\textbf {\bibinfo {volume} {109}},\ \bibinfo
  {pages} {224315} (\bibinfo {year} {2024})}\BibitemShut {NoStop}%
\bibitem [{\citenamefont {Song}\ \emph {et~al.}(2017)\citenamefont {Song},
  \citenamefont {Fang},\ and\ \citenamefont {Fang}}]{PhysRevLett.119.246402}%
  \BibitemOpen
  \bibfield  {author} {\bibinfo {author} {\bibfnamefont {Z.}~\bibnamefont
  {Song}}, \bibinfo {author} {\bibfnamefont {Z.}~\bibnamefont {Fang}},\ and\
  \bibinfo {author} {\bibfnamefont {C.}~\bibnamefont {Fang}},\ }\bibfield
  {title} {\bibinfo {title} {$(d\ensuremath{-}2)$-dimensional edge states of
  rotation symmetry protected topological states},\ }\href
  {https://doi.org/10.1103/PhysRevLett.119.246402} {\bibfield  {journal}
  {\bibinfo  {journal} {Phys. Rev. Lett.}\ }\textbf {\bibinfo {volume} {119}},\
  \bibinfo {pages} {246402} (\bibinfo {year} {2017})}\BibitemShut {NoStop}%
\bibitem [{\citenamefont {Langbehn}\ \emph {et~al.}(2017)\citenamefont
  {Langbehn}, \citenamefont {Peng}, \citenamefont {Trifunovic}, \citenamefont
  {von Oppen},\ and\ \citenamefont {Brouwer}}]{PhysRevLett.119.246401}%
  \BibitemOpen
  \bibfield  {author} {\bibinfo {author} {\bibfnamefont {J.}~\bibnamefont
  {Langbehn}}, \bibinfo {author} {\bibfnamefont {Y.}~\bibnamefont {Peng}},
  \bibinfo {author} {\bibfnamefont {L.}~\bibnamefont {Trifunovic}}, \bibinfo
  {author} {\bibfnamefont {F.}~\bibnamefont {von Oppen}},\ and\ \bibinfo
  {author} {\bibfnamefont {P.~W.}\ \bibnamefont {Brouwer}},\ }\bibfield
  {title} {\bibinfo {title} {Reflection-symmetric second-order topological
  insulators and superconductors},\ }\href
  {https://doi.org/10.1103/PhysRevLett.119.246401} {\bibfield  {journal}
  {\bibinfo  {journal} {Phys. Rev. Lett.}\ }\textbf {\bibinfo {volume} {119}},\
  \bibinfo {pages} {246401} (\bibinfo {year} {2017})}\BibitemShut {NoStop}%
\bibitem [{\citenamefont {Schindler}\ \emph {et~al.}(2018)\citenamefont
  {Schindler}, \citenamefont {Cook}, \citenamefont {Vergniory}, \citenamefont
  {Wang}, \citenamefont {Parkin}, \citenamefont {Bernevig},\ and\ \citenamefont
  {Neupert}}]{Schindler2018-uw}%
  \BibitemOpen
  \bibfield  {author} {\bibinfo {author} {\bibfnamefont {F.}~\bibnamefont
  {Schindler}}, \bibinfo {author} {\bibfnamefont {A.~M.}\ \bibnamefont {Cook}},
  \bibinfo {author} {\bibfnamefont {M.~G.}\ \bibnamefont {Vergniory}}, \bibinfo
  {author} {\bibfnamefont {Z.}~\bibnamefont {Wang}}, \bibinfo {author}
  {\bibfnamefont {S.~S.~P.}\ \bibnamefont {Parkin}}, \bibinfo {author}
  {\bibfnamefont {B.~A.}\ \bibnamefont {Bernevig}},\ and\ \bibinfo {author}
  {\bibfnamefont {T.}~\bibnamefont {Neupert}},\ }\bibfield  {title} {\bibinfo
  {title} {Higher-order topological insulators},\ }\href
  {https://doi.org/10.1126/sciadv.aat0346} {\bibfield  {journal} {\bibinfo
  {journal} {Sci. Adv.}\ }\textbf {\bibinfo {volume} {4}},\ \bibinfo {pages}
  {eaat0346} (\bibinfo {year} {2018})}\BibitemShut {NoStop}%
\bibitem [{\citenamefont {Serra-Garcia}\ \emph {et~al.}(2018)\citenamefont
  {Serra-Garcia}, \citenamefont {Peri}, \citenamefont {S{\"u}sstrunk},
  \citenamefont {Bilal}, \citenamefont {Larsen}, \citenamefont {Villanueva},\
  and\ \citenamefont {Huber}}]{serra2018observation}%
  \BibitemOpen
  \bibfield  {author} {\bibinfo {author} {\bibfnamefont {M.}~\bibnamefont
  {Serra-Garcia}}, \bibinfo {author} {\bibfnamefont {V.}~\bibnamefont {Peri}},
  \bibinfo {author} {\bibfnamefont {R.}~\bibnamefont {S{\"u}sstrunk}}, \bibinfo
  {author} {\bibfnamefont {O.~R.}\ \bibnamefont {Bilal}}, \bibinfo {author}
  {\bibfnamefont {T.}~\bibnamefont {Larsen}}, \bibinfo {author} {\bibfnamefont
  {L.~G.}\ \bibnamefont {Villanueva}},\ and\ \bibinfo {author} {\bibfnamefont
  {S.~D.}\ \bibnamefont {Huber}},\ }\bibfield  {title} {\bibinfo {title}
  {Observation of a phononic quadrupole topological insulator},\ }\href
  {https://www.nature.com/articles/nature25156} {\bibfield  {journal} {\bibinfo
   {journal} {Nature}\ }\textbf {\bibinfo {volume} {555}},\ \bibinfo {pages}
  {342} (\bibinfo {year} {2018})}\BibitemShut {NoStop}%
\bibitem [{\citenamefont {Peterson}\ \emph {et~al.}(2018)\citenamefont
  {Peterson}, \citenamefont {Benalcazar}, \citenamefont {Hughes},\ and\
  \citenamefont {Bahl}}]{peterson2018quantized}%
  \BibitemOpen
  \bibfield  {author} {\bibinfo {author} {\bibfnamefont {C.~W.}\ \bibnamefont
  {Peterson}}, \bibinfo {author} {\bibfnamefont {W.~A.}\ \bibnamefont
  {Benalcazar}}, \bibinfo {author} {\bibfnamefont {T.~L.}\ \bibnamefont
  {Hughes}},\ and\ \bibinfo {author} {\bibfnamefont {G.}~\bibnamefont {Bahl}},\
  }\bibfield  {title} {\bibinfo {title} {A quantized microwave quadrupole
  insulator with topologically protected corner states},\ }\href
  {https://www.nature.com/articles/nature25777} {\bibfield  {journal} {\bibinfo
   {journal} {Nature}\ }\textbf {\bibinfo {volume} {555}},\ \bibinfo {pages}
  {346} (\bibinfo {year} {2018})}\BibitemShut {NoStop}%
\bibitem [{\citenamefont {Zhang}\ \emph {et~al.}(2021)\citenamefont {Zhang},
  \citenamefont {Zou}, \citenamefont {Pei}, \citenamefont {He}, \citenamefont
  {Bao}, \citenamefont {Sun},\ and\ \citenamefont
  {Zhang}}]{PhysRevLett.126.146802}%
  \BibitemOpen
  \bibfield  {author} {\bibinfo {author} {\bibfnamefont {W.}~\bibnamefont
  {Zhang}}, \bibinfo {author} {\bibfnamefont {D.}~\bibnamefont {Zou}}, \bibinfo
  {author} {\bibfnamefont {Q.}~\bibnamefont {Pei}}, \bibinfo {author}
  {\bibfnamefont {W.}~\bibnamefont {He}}, \bibinfo {author} {\bibfnamefont
  {J.}~\bibnamefont {Bao}}, \bibinfo {author} {\bibfnamefont {H.}~\bibnamefont
  {Sun}},\ and\ \bibinfo {author} {\bibfnamefont {X.}~\bibnamefont {Zhang}},\
  }\bibfield  {title} {\bibinfo {title} {Experimental observation of
  higher-order topological anderson insulators},\ }\href
  {https://doi.org/10.1103/PhysRevLett.126.146802} {\bibfield  {journal}
  {\bibinfo  {journal} {Phys. Rev. Lett.}\ }\textbf {\bibinfo {volume} {126}},\
  \bibinfo {pages} {146802} (\bibinfo {year} {2021})}\BibitemShut {NoStop}%
\bibitem [{\citenamefont {Moore}\ and\ \citenamefont
  {Read}(1991)}]{MOORE1991362}%
  \BibitemOpen
  \bibfield  {author} {\bibinfo {author} {\bibfnamefont {G.}~\bibnamefont
  {Moore}}\ and\ \bibinfo {author} {\bibfnamefont {N.}~\bibnamefont {Read}},\
  }\bibfield  {title} {\bibinfo {title} {Nonabelions in the fractional quantum
  hall effect},\ }\href
  {https://doi.org/https://doi.org/10.1016/0550-3213(91)90407-O} {\bibfield
  {journal} {\bibinfo  {journal} {Nucl. Phys. B.}\ }\textbf {\bibinfo {volume}
  {360}},\ \bibinfo {pages} {362} (\bibinfo {year} {1991})}\BibitemShut
  {NoStop}%
\bibitem [{\citenamefont {Nayak}\ \emph
  {et~al.}(2008{\natexlab{b}})\citenamefont {Nayak}, \citenamefont {Simon},
  \citenamefont {Stern}, \citenamefont {Freedman},\ and\ \citenamefont
  {Das~Sarma}}]{RevModPhys.80.1083}%
  \BibitemOpen
  \bibfield  {author} {\bibinfo {author} {\bibfnamefont {C.}~\bibnamefont
  {Nayak}}, \bibinfo {author} {\bibfnamefont {S.~H.}\ \bibnamefont {Simon}},
  \bibinfo {author} {\bibfnamefont {A.}~\bibnamefont {Stern}}, \bibinfo
  {author} {\bibfnamefont {M.}~\bibnamefont {Freedman}},\ and\ \bibinfo
  {author} {\bibfnamefont {S.}~\bibnamefont {Das~Sarma}},\ }\bibfield  {title}
  {\bibinfo {title} {Non-abelian anyons and topological quantum computation},\
  }\href {https://doi.org/10.1103/RevModPhys.80.1083} {\bibfield  {journal}
  {\bibinfo  {journal} {Rev. Mod. Phys.}\ }\textbf {\bibinfo {volume} {80}},\
  \bibinfo {pages} {1083} (\bibinfo {year} {2008}{\natexlab{b}})}\BibitemShut
  {NoStop}%
\bibitem [{\citenamefont {Guo}\ \emph {et~al.}(2021)\citenamefont {Guo},
  \citenamefont {Jiang}, \citenamefont {Zhang}, \citenamefont {Zhang},
  \citenamefont {Zhang}, \citenamefont {Yang}, \citenamefont {Zhang},\ and\
  \citenamefont {Chan}}]{guo2021experimental}%
  \BibitemOpen
  \bibfield  {author} {\bibinfo {author} {\bibfnamefont {Q.}~\bibnamefont
  {Guo}}, \bibinfo {author} {\bibfnamefont {T.}~\bibnamefont {Jiang}}, \bibinfo
  {author} {\bibfnamefont {R.-Y.}\ \bibnamefont {Zhang}}, \bibinfo {author}
  {\bibfnamefont {L.}~\bibnamefont {Zhang}}, \bibinfo {author} {\bibfnamefont
  {Z.-Q.}\ \bibnamefont {Zhang}}, \bibinfo {author} {\bibfnamefont
  {B.}~\bibnamefont {Yang}}, \bibinfo {author} {\bibfnamefont {S.}~\bibnamefont
  {Zhang}},\ and\ \bibinfo {author} {\bibfnamefont {C.~T.}\ \bibnamefont
  {Chan}},\ }\bibfield  {title} {\bibinfo {title} {Experimental observation of
  non-abelian topological charges and edge states},\ }\href
  {https://www.nature.com/articles/s41586-021-03521-3} {\bibfield  {journal}
  {\bibinfo  {journal} {Nature}\ }\textbf {\bibinfo {volume} {594}},\ \bibinfo
  {pages} {195} (\bibinfo {year} {2021})}\BibitemShut {NoStop}%
\bibitem [{\citenamefont {Sakaguchi}\ \emph {et~al.}(2016)\citenamefont
  {Sakaguchi}, \citenamefont {Sherman},\ and\ \citenamefont
  {Malomed}}]{PhysRevE.94.032202}%
  \BibitemOpen
  \bibfield  {author} {\bibinfo {author} {\bibfnamefont {H.}~\bibnamefont
  {Sakaguchi}}, \bibinfo {author} {\bibfnamefont {E.~Y.}\ \bibnamefont
  {Sherman}},\ and\ \bibinfo {author} {\bibfnamefont {B.~A.}\ \bibnamefont
  {Malomed}},\ }\bibfield  {title} {\bibinfo {title} {Vortex solitons in
  two-dimensional spin-orbit coupled {Bose-Einstein} condensates: Effects of
  the {Rashba-Dresselhaus} coupling and {Zeeman} splitting},\ }\href
  {https://doi.org/10.1103/PhysRevE.94.032202} {\bibfield  {journal} {\bibinfo
  {journal} {Phys. Rev. E}\ }\textbf {\bibinfo {volume} {94}},\ \bibinfo
  {pages} {032202} (\bibinfo {year} {2016})}\BibitemShut {NoStop}%
\bibitem [{\citenamefont {Pang}\ \emph {et~al.}(2018)\citenamefont {Pang},
  \citenamefont {Deng}, \citenamefont {Liu}, \citenamefont {Xu},\ and\
  \citenamefont {Li}}]{Pang2018-aa}%
  \BibitemOpen
  \bibfield  {author} {\bibinfo {author} {\bibfnamefont {W.}~\bibnamefont
  {Pang}}, \bibinfo {author} {\bibfnamefont {H.}~\bibnamefont {Deng}}, \bibinfo
  {author} {\bibfnamefont {B.}~\bibnamefont {Liu}}, \bibinfo {author}
  {\bibfnamefont {J.}~\bibnamefont {Xu}},\ and\ \bibinfo {author}
  {\bibfnamefont {Y.}~\bibnamefont {Li}},\ }\bibfield  {title} {\bibinfo
  {title} {Two-dimensional vortex solitons in spin-orbit-coupled dipolar
  {Bose-Einstein} condensates},\ }\href
  {https://www.mdpi.com/2076-3417/8/10/1771} {\bibfield  {journal} {\bibinfo
  {journal} {Appl. Sci.}\ }\textbf {\bibinfo {volume} {8}},\ \bibinfo {pages}
  {1771} (\bibinfo {year} {2018})}\BibitemShut {NoStop}%
\bibitem [{\citenamefont {Xu}\ \emph {et~al.}(2024)\citenamefont {Xu},
  \citenamefont {Wu}, \citenamefont {Hu}, \citenamefont {He}, \citenamefont
  {Zhao},\ and\ \citenamefont {Fan}}]{XU2024115043}%
  \BibitemOpen
  \bibfield  {author} {\bibinfo {author} {\bibfnamefont {S.-L.}\ \bibnamefont
  {Xu}}, \bibinfo {author} {\bibfnamefont {T.}~\bibnamefont {Wu}}, \bibinfo
  {author} {\bibfnamefont {H.-J.}\ \bibnamefont {Hu}}, \bibinfo {author}
  {\bibfnamefont {J.-R.}\ \bibnamefont {He}}, \bibinfo {author} {\bibfnamefont
  {Y.}~\bibnamefont {Zhao}},\ and\ \bibinfo {author} {\bibfnamefont
  {Z.}~\bibnamefont {Fan}},\ }\bibfield  {title} {\bibinfo {title} {Vortex
  solitons in {Rydberg-excited} {Bose-Einstein} condensates with rotating
  {PT-symmetric} azimuthal potentials},\ }\href
  {https://doi.org/https://doi.org/10.1016/j.chaos.2024.115043} {\bibfield
  {journal} {\bibinfo  {journal} {Chaos, Solitons \& Fractals}\ }\textbf
  {\bibinfo {volume} {184}},\ \bibinfo {pages} {115043} (\bibinfo {year}
  {2024})}\BibitemShut {NoStop}%
\bibitem [{\citenamefont {Perapechka}\ \emph {et~al.}(2018)\citenamefont
  {Perapechka}, \citenamefont {Sawado},\ and\ \citenamefont
  {Shnir}}]{perapechka2018soliton}%
  \BibitemOpen
  \bibfield  {author} {\bibinfo {author} {\bibfnamefont {I.}~\bibnamefont
  {Perapechka}}, \bibinfo {author} {\bibfnamefont {N.}~\bibnamefont {Sawado}},\
  and\ \bibinfo {author} {\bibfnamefont {Y.}~\bibnamefont {Shnir}},\ }\bibfield
   {title} {\bibinfo {title} {Soliton solutions of the fermion-skyrmion system
  in (2+1) dimensions},\ }\href
  {https://link.springer.com/article/10.1007/JHEP10(2018)081} {\bibfield
  {journal} {\bibinfo  {journal} {J. High Energy Phys.}\ }\textbf {\bibinfo
  {volume} {2018}}\bibinfo  {number} { (10)},\ \bibinfo {pages}
  {1}}\BibitemShut {NoStop}%
\bibitem [{\citenamefont {Liu}\ \emph {et~al.}(2013)\citenamefont {Liu},
  \citenamefont {Zhang},\ and\ \citenamefont {Yang}}]{LIU20133300}%
  \BibitemOpen
\bibfield  {number} {  }\bibfield  {author} {\bibinfo {author} {\bibfnamefont
  {Y.-K.}\ \bibnamefont {Liu}}, \bibinfo {author} {\bibfnamefont
  {C.}~\bibnamefont {Zhang}},\ and\ \bibinfo {author} {\bibfnamefont {S.-J.}\
  \bibnamefont {Yang}},\ }\bibfield  {title} {\bibinfo {title} {3d skyrmion and
  knot in two-component {Bose–Einstein} condensates},\ }\href
  {https://doi.org/https://doi.org/10.1016/j.physleta.2013.10.025} {\bibfield
  {journal} {\bibinfo  {journal} {Phys. Lett. A}\ }\textbf {\bibinfo {volume}
  {377}},\ \bibinfo {pages} {3300} (\bibinfo {year} {2013})}\BibitemShut
  {NoStop}%
\end{thebibliography}

%

\end{document}